\documentstyle[11pt]{article}

\newcommand{\ncm}{\newcommand}
\ncm{\rncm}{\renewcommand}
\rncm{\sec}{\setc{0}\section}
\ncm{\bsn}{\bigskip\noindent}
\ncm{\beq}{\begin{equation}}
\ncm{\eeq}{\end{equation}}
\ncm{\bea}{\begin{eqnarray}}
\ncm{\beanon}{\begin{eqnarray*}}
\ncm{\eea}{\end{eqnarray}}
\ncm{\eeanon}{\end{eqnarray*}}
\ncm{\fns}{\footnotesize}

\rncm{\theequation}{\thesection.\arabic{equation}}
\ncm{\setc}[1]{\setcounter{equation}{#1}}
\newcounter{eqnr}

\newenvironment{eqnarrayabc}{\stepcounter{equation}
  \setcounter{eqnr}{\value{equation}}\setc{0}
  \rncm{\theequation}{\thesection.\arabic{eqnr}\alph{equation}}
  \begin{eqnarray}}{\end{eqnarray}\setc{\value{eqnr}}}
\ncm{\bealph}{\begin{eqnarrayabc}}
\ncm{\eealph}{\end{eqnarrayabc}}
\ncm{\bit}{\begin{itemize}}
\ncm{\eit}{\end{itemize}}
\ncm{\eqboxabc}[3]{\newline\parbox[t]{1.5cm}{#1}\hfill
  \parbox[b]{12cm}{\begin{eqnarray*} #3\end{eqnarray*}}\hfill
   \parbox[b]{1.5cm}{\vspace{-0.0cm}
  \begin{eqnarrayabc}#2\end{eqnarrayabc}}\newline}

\ncm{\eqbox}[2]{\newline\parbox{1.5cm}{#1}\hfill
  \parbox{12cm}{\beanon #2\eeanon}\hfill
  \parbox{1cm}{\bea\eea}\newline}

\newcommand{\eqa}[2]{\begin{flushright}
 $\begin{array}{rclr}&&&\hspace{#1}\\[-1em]
 #2 \end{array}$
 \end{flushright}}

\ncm{\half}{{1\over 2}}
\def\lcros{\raise1.5pt\hbox{$\scriptstyle\triangleright$}\!
           \raise1.9pt\hbox{$\scriptscriptstyle < \,$}}

\def\lg{{{\lower1.5ex\hbox{$< $}}\atop{\raise1.5ex\hbox{$> $}}}}
\def\truesupset{{\lower5pt\hbox{$\scriptstyle\supset$}\atop
 \raise5pt\hbox{$\scriptscriptstyle\not=$}}}
\def\truesubset{{\lower5pt\hbox{$\scriptstyle\subset$}\atop
 \raise5pt\hbox{$\scriptscriptstyle\not=$}}}
\def\E{\hbox{End}\,}
\def\End{\hbox{End}\,}

\def\Aut{\hbox{Aut}\,}
\def\Hom{\hbox{Hom}\,}

\def\Int{\hbox{Int}\,}
\def\Im{\hbox{Im}\,}

\def\id{\hbox{id}\,}
\def\idA{\hbox{id}_{\A}}
\def\idG{\hbox{id}_{\G}}
\def\Del{\Delta_{\D}}
\def\Delop{\Delta_{\D}^{op}}

\def\reven{\rho_{2i,2i+1}}
\def\rodd{\rho_{2i-1,2i}}
\def\Ad{\hbox{Ad}\,}
\def\Amp{{\bf Amp\,}}

\def\Rep{{\bf Rep\,}}
\def\F{{\cal F}}

\def\BB{{\bf B}}

\def\W{{\cal W}}
\def\OA{{\cal A}}
\def\I{{\cal I}}

\def\one{{\bf 1}}
\def\onne{{\thinmuskip=5.5mu 1\!1\thinmuskip=3mu}}
\def\o{\otimes}

\def\x{\times}

\def\mur{{\buildrel\lower6pt\hbox{$\scriptstyle
         \rightarrow$}\over\mu}}
\def\mul{{\buildrel\lower6pt\hbox{$\scriptstyle
         \leftarrow$}\over\mu}}

\def\cros{\raise1.9pt\hbox{$\scriptscriptstyle
          > $}\!\raise1.5pt\hbox{$\scriptstyle\triangleleft\,$}}
\def\RR{{I\!\!R}}
\def\CC{\,{\raise1.5pt\hbox{$\scriptscriptstyle |$}
        \thinmuskip=4mu \!\!C\thinmuskip=3mu}}
\def\ZZ{{Z\!\!\!Z}}
\def\NN{{\thinmuskip = 5.5mu I\!N\thinmuskip = 3mu}}
\def\N{\NN}
\def\Z{\ZZ}

\def\r{\rho}

\def\W{{\cal W}}
\def\L{{\cal L}}

\def\bra{\langle}
\def\ket{\rangle}
\def\qed{\hfill {\it Q.e.d.}\\}

\newcommand{\rt}{&\hspace{-1em}}
\newcommand{\lt}{\hspace{-1em}&}
\newcommand{\A}{{\cal A}}
\newcommand{\B}{{\cal B}}
\newcommand{\G}{{\cal G}}
\renewcommand{\H}{{\cal H}}
\newcommand{\C}{{\cal C}}
\newcommand{\D}{{\cal D}}
\newcommand{\T}{{\cal T}}
\newcommand{\1}{{\bf 1}}

\ncm{\ea}{\end{array}}
\ncm{\ba}{\begin{array}}

\begin{document}

\title{\bf Quantum Chains of Hopf Algebras\\
  \bf with Quantum Double Cosymmetry}
\protect\author{{\sc Florian Nill $^1$}
\\
{\small Institut f\"ur Theoretische Physik,
FU-Berlin, Arnimallee 14, D-14195 Berlin}
\and
{\sc Korn\'el Szlach\'anyi\ $^2$}
\\
{\small Central Research Institute for Physics
H-1525 Budapest 114, P.O.B. 49, Hungary}}

\date{August 1995}

\maketitle

\footnotetext[1]{Supported by DFG, SFB 288 "Differentialgeometrie und
   Quantenphysik"; email: nill@physik.fu-berlin.de}
\footnotetext[2]{Supported by the Hungarian Scient. Res. Fund,
  OTKA--1815;
email: szlach@rmki.kfki.hu}

\vspace{-9cm}
\begin{flushright}
Revised Aug. 1997\\
Published slightly shortened in\\
 Comm. Math. Phys. {\bf 187} (1997) 159 - 200
\end{flushright}

\vspace{6cm}

\begin{abstract}

Given a finite dimensional $C^*$-Hopf algebra $H$ and its dual
$\hat H$ we construct the infinite crossed product
$\A=\dots\cros H\cros\hat H\cros H\cros\dots$
and study its superselection sectors in the framework of algebraic
quantum field theory.
$\A$ is the observable algebra
of a generalized quantum spin chain with $H$-order and $\hat H$-disorder
symmetries, where by a duality transformation the role of order
and disorder may also appear interchanged.
If $H=\CC G$ is a group algebra then $\A$ becomes an ordinary
$G$-spin model. We classify
all DHR-sectors of $\A$ --- relative to some Haag dual vacuum
representation --- and prove that their symmetry is described by the
Drinfeld double $\D(H)$.
To achieve this we construct {\em localized coactions }
$\r:\A\to\A\o\D(H)$ and use a certain compressibility property
to prove that they are {\em universal amplimorphisms} on $\A$.
In this way the double $\D(H)$ can be recovered from the
observable algebra $\A$ as a {\em universal cosymmetry}.

\end{abstract}

\footnotesize\tableofcontents

\normalsize

\sec{Introduction and Summary of Results}

Quantum chains considered as models of $1+1$-dimensional quantum
field theory exhibit many interesting features that are either
impossible or unknown in higher ($2+1$ or $3+1$) dimensions.
These features include integrability on the one hand and the
emergence of braid group statistics and quantum symmetry on the
other hand.
In this paper we study the second class of phenomena by looking
at Hopf spin models as a general class of
quantum chains where the quantum symmetry and braid statistics
of superselection sectors turns out to be described by
Drinfeld's ``quantum double" $\D(H)$ of the underlying Hopf
algebra $H$.

Quantum chains on which a quantum group acts are well known for
some time; for example the XXZ-chain with the action of $sl(2)_q$
[P,PS] or the lattice Kac--Moody algebras of [AFSV,AFS,Fa,FG].
For a recent paper on the general action of quantum
groups on ultralocal quantum chains see [FNW].
However the discovery that --- at least for non-integer
statistical dimensions ---quantum symmetries are described
by truncated quasi-Hopf algebras
[MS1-2,S] presents new difficulties to this approach.
In fact, in such a scenario the ``field algebras'' are
non-associative and do not obey commutation relations with
$c$-number coefficients, both properties being tacidly
assumed in any ``decent" quantum chain.

In continuum theories quantum double symmetries have also been
realized in orbifold models [DPR] and in integrable models (see [BL]
for a review). For a recent axiomatic approach within the scheme of
algebraic quantum field theory see [M]. In contrast with our approach,
in these papers the fields transforming non-trivially under an
``order'' symmetry $H$ are already assumed to be given in the theory
from the beginnig,
and the task reduces to constructing the disorder fields transforming
under the dual $\hat H$.

Here we stress the point of view that an unbiased
approach to reveal the quantum symmetry of a model must be based
only on the knowledge of the quantum group invariant operators
(the "observables") that obey local commutation relations. This is
the approach of algebraic quantum field theory (AQFT) [H].
The importance of the algebraic method, in
particular the DHR theory of superselection sectors [DHR],
in low dimensional QFT has been realized by many authors (see
[FRS,BMT,Fr\"oGab,F,R] and many others).

The implementation
of the DHR theory to quantum chains has been carried out at first
for the case of $G$-spin models in [SzV]. These models have an
order-disorder type of quantum symmetry given by the double
$\D(G)$ of a finite group $G$ which generalizes the $Z(2)\times
Z(2)$ symmetry of the lattice Ising model. Since the disorder part
of the double (i.e. the function algebra $\C(G)$) is always
Abelian, $G$-spin models cannot be selfdual in the Kramers-Wannier
sense, unless the group is Abelian. Non-Abelian Kramers-Wannier
duality can therefore be expected only in a larger class of
models.

Here we shall investigate the following generalization of
$G$-spin models. On each lattice site there is a copy of a finite
dimensional $C^*$-Hopf algebra $H$ and on each link there is a
copy of its dual $\hat H$. Non-trivial commutation relations are
postulated only between neighbor links and sites where $H$ and
$\hat H$ act on each other in the "natural way", so as the link-site and the
site-link algebras to form the crossed products
${\cal W}(\hat H)\equiv\hat H\cros H$ and ${\cal W}(H)\equiv
H\cros\hat H$ ("Weyl algebras" in the terminology of [N]).
 The two-sided infinite crossed
product $\dots\cros H\cros\hat H\cros H\cros\hat H\cros\dots$
defines the observable algebra $\A$ of the Hopf spin model. Its
superselection sectors (more precisely those that correspond to charges
localized within a finite interval $I$, the so called DHR sectors)
can be created by localized
amplimorphisms $\mu\colon\A\to\A\o\End V$ with $V$ denoting
some finite dimensional Hilbert space.
The category of localized amplimorphisms
$\Amp\A$ plays the same role in locally finite dimensional theories
as the category ${\bf End\,}\A$ of localized
endomorphisms in continuum theories.
The symmetry of the superselection sectors can be revealed by finding
the ``quantum group'' $\G$,
the representation category of which is equivalent
to $\Amp\A$. In our model we find that $\G$ is
the Drinfeld double (also called the quantum double) $\D(H)$
of $H$.

Finding all endomorphisms or all amplimorphisms of a given observable
algebra $\A$ can be a very difficult problem in general. In the Hopf spin
model $\A$ possesses a property we call {\it complete compressibility},
which allows us to do so. Namely if $\mu$ is an amplimorphism creating
some charge on an arbirary large but finite interval then there
exists an amplimorphism $\nu$ creating the same charge (i.e. $\nu$ is
equivalent to $\mu$, written $\nu\sim\mu$) but within an interval $I$
of length 2 (i.e. $I$ consists of a neighbouring site--link pair).
Therefore the problem of finding all
DHR-sectors of the Hopf spin model is reduced to a finite
dimensional problem, namely to find all amplimorphisms localized
within an interval of length 2. In this way we have proven that
{\em all} DHR-sectors of $\A$ can be classified by representations of
the Drinfeld double.

An important role in this reconstruction
is played by the so-called {\it universal} amplimorphisms in $\Amp\A$.
 These are amplimorphisms
$\rho\colon\A\to\A\o\G$ where $\G$ is an appropriate (in our
approach finite dimensional) ``quantum symmetry" $C^*$-algebra
such that for any other amplimorphism $\mu$ in $\Amp\A$
there exists a representation $\beta_{\mu}$ of
$\G$ such that
$\mu\sim(\idA\o\beta_{\mu})\circ\rho$. Moreover,
the correspondence $\mu\leftrightarrow\beta_\mu$ has to be one-to-one
on equivalence classes.
We prove that complete compressibility implies that universal
amplimorphisms $\r$ can be chosen to provide {\em coactions} of
$\G$ on $\A$, i.e. there exists a coassociative unital
coproduct $\Delta:\G\to\G\o\G$ and a counit
$\varepsilon:\G\to\CC$ such that
\bealph
(\rho\o\idG)\circ\rho\ &=&\ (\idA\o\Delta)\circ\rho\\
(\idA\o\varepsilon)\circ\rho\ &=&\ \idA
\eealph
Moreover, $\Delta$ and $\varepsilon$ are uniquely determined by
$\r$. Thus $\G$ becomes a $C^*$-Hopf algebra which we call
a {\em universal cosymmetry} of $\A$.
$\G$ will in fact be quasitriangular
with $R$-matrix determined by the
statistics operator of $\rho$
\beq
\epsilon(\rho,\rho)\ =\ \onne_\A\o P^{12}R
\eeq
where $R\in\G\o\G$ and where $P^{12}$ is the usual permutation. The antipode
$S$ of $\G$ can be recovered by studying conjugate objects $\bar\rho$
and intertwiners $\rho\times\bar\rho\to\idA$.
In this type of models
the statistical dimensions $d_r$ of the irreducible components
$\rho_r$ of $\rho$ are integers: they coincide with the dimensions
of the corresponding irreducible representation $D_r$ of $\G$.
The statistics phases can be obtained from the universal balancing
element $s=S(R_2)R_1\in\hbox{Center}\,\G$ evaluated in
the representations $D_r$.
For the Hopf spin model this scenario can be verified and
calculated explicitely with $\G=\D(H)$.

\smallskip
We emphasize that being a universal cosymmetry $\G$ is uniquely
determined as a $C^*$-algebra together with a distinguished 1-dimensional
representation $\varepsilon$. The dimensions of irreps of $\G$
coincide with the statistical dimensions of the associated sectors of
$\A$, $n_r=d_r$, the latter being integer valued.
This has to be contrasted with the approaches based on truncated
(quasi) Hopf algebras [MS2,S,FGV], where the $n_r$'s are only constrained by an
inequality involving the fusion matrices.
In this sense our construction parallels the Doplicher-Roberts
approach [DR1,2], where $\G$ would be a group algebra. 

However, it is important to note that given $\Amp\A\sim\Rep\G$ as
braided rigid $C^*$-tensor categories does not fix the coproduct on
$\G$ uniquely, even not in the case of group algebras. More precisely,
the quasitriangular Hopf algebra structure on
$\G$ can be recovered only up to a twisting by a 2-cocycle:
If $u\in\G\o\G$ is a 2-cocycle, i.e. a unitary satisfying
\bealph
(u\o\one)\cdot(\Delta\o\id)(u)\ &=&\ (\one\o
u)\cdot(\id\o\Delta)(u)\,,\\
(\varepsilon\o\id)(u)\ &=&\ (\id\o\varepsilon)(u)\ =\ \one
\eealph
then the twisted quasitriangular Hopf algebra with data
\beanon
\Delta'\ &=&\ \Ad u\circ\Delta\\
\varepsilon'\ &=&\ \varepsilon\\
S'\ &=&\ \Ad q\circ S\qquad q:=u_1S(u_2)\\
R'\ &=&\ u^{op}Ru^*\\
\eeanon
is as good for a (co-)symmetry as the original one. In fact, we prove in
Section 3.5 that (up to transformations by $\sigma\in\Aut(\G,\varepsilon)$)
any universal coaction $(\rho',\Delta')$ is
equivalent to a fixed one $(\r,\Delta)$
by an isometric intertwiner $U\in\A\o\G$ satisfying a {\it  twisted
cocycle condition}
\bealph
U\rho(A)\ &=&\ \rho'(A)U,\quad A\in\A,\\
(U\o\one)\cdot(\rho\o\idG)(U)\ &=&
\ (\onne\o u)\cdot(\idA\o\Delta)(U)\,,\\
(\idA\o\varepsilon)(U)\ &=&\ \onne
\eealph
implying the identities (1.3) for $u$.
In the Hopf spin model we also have the reverse statement, i.e.
for all 2-cocycles $u$ there is a unitary $U\in\A\o\G$
and a universal coaction
$\rho'$ satisfying (1.4) and therefore (1.1) with $\Delta'$ instead of
$\Delta$.
We point out that (1.4) is a generalization of the usual notion
of cocycle equivalence for coactions where one requires
$u=\one\o\one$ [Ta,NaTa,BaSk,E]. To our knowledge, in the DR-approach
[DR1,2] this possibility of twisting has not been considered, since
there it would seem ``unnatural'' to deviate from the standard
coproduct on a group algebra.

\medskip
This paper is an extended version of the first part of [NSz1].
In a forthcoming paper we will show [NSz3] that any universal
coaction $\r$ on $\A$ gives rise to a family of complete irreducible
field algebra extensions $\F\supset\A$ and that all field
algebra extensions of $\A$ arise in this way.
Moreover, equivalence
classes of complete irreducible
field algebra extensions are in one-to-one
correspondence with cohomology classes of 2-cocycles $u\in\G\o\G$.
The Hopf algebra $\G$ will act as a global gauge symmetry on
all $\F$'s such that $\A\subset\F$ is precisely the $\G$-invariant subalgebra.
Inequivalent field algebras will be shown to be related by Klein
transformations involving symmetry operators $Q(X),\ X\in\G$.

\medskip
The above type of
reconstruction of the quasitriangular Hopf algebra $\G$ is
a special case of the generalized Tannaka-Krein theorem
[U,Maj2]. Namely, any faithful functor
$F\colon{\cal C}\to Vec$ from strict monoidal braided rigid
$C^*$-categories to the category of finite dimensional
vector spaces factorizes as $F=f\circ\Phi$ to the forgetful
functor $f$ and to an equivalence $\Phi$ of ${\cal C}$ with the
representation category $\Rep\G$ of a quasitriangular $C^*$-Hopf
algebra $\G$. In our case ${\cal C}$ is the category $\Amp\A$ of
amplimorphisms of the observable algebra $\A$. The functor $F$ to
the vector spaces is given naturally by associating to the
amplimorphism $\mu\colon\A\to\A\o\End V$ the vector space $V$.
Although the vector spaces $V$ cannot be seen by only looking at the
abstract category $\Amp\A$, they are "inherently" determined by the
amplimorphisms and therefore by the observable algebra itself.
In this respect using amplimorphisms one goes somewhat beyond the
Tannaka-Krein theorem and approaches a Doplicher-Roberts [DR] type of
reconstruction.

\medskip
We now describe the plan of this paper.

In Section 2.1 we define our model using abstract relations as
well as concrete realizations on Hilbert spaces associated to
finite lattice intervals. We also discuss duality
transformations and the appearence of the Drinfeld double as an
order-disorder symmetry. In Section 2.2 we present the notion
of a {\em quantum Gibbs system} on $\A$ and use this to prove
(algebraic) Haag duality of our model.

In Section 3 we start with reviewing the category of
amplimorphisms $\Amp\A$ in Section 3.1 and introduce
{\em localized cosymmetries} $\r:\A\to\A\o\G$
as special kinds of amplimorphisms in Section 3.2.
In Section 3.3 we specialize to {\em effective cosymmetries} and
show that $\Amp\A\sim\Rep\G$ provided $\G$ is also {\em universal}.
In Section 3.4 we introduce and investigate the notion of {\em complete
compressibility} to guarantee the existence of universal
cosymmetries. In Section 3.5 we prove that universal
cosymmetries are unique up to (twisted) cocycle equivalences.
In Section 3.6 we discuss two notions of translation covariance
for localized cosymmetries and relate these to the existence of
a {\em coherently translation covariant} structure in $\Amp\A$
as introduced for the case of endomorphisms in [DR1].

In Section 4 we apply the general theory to our Hopf spin model.
In Section 4.1 we construct localized and strictly translation
covariant effective coactions
$\r_I:\A\to\A\o\D(H)$ of the Drinfeld double for any interval
$I$ of length two and in Section 4.2 we prove that all these
coactions are actually universal in $\Amp\A$.

\bsn
{\em Remarks added in the revised version:}\\
 Meanwhile (i.e. 9 months after releasing our first preprint),
the notion of a localized coaction has also been taken up
in a paper by Alekseev, Faddeev, Fr\"ohlich and Schomerus [AFFS]
without referring to our work.
In fact, the lattice
current algebra studied by [AFFS] (which is an extension of
[AFSV,AFS,FG]) has meanwhile been realized by one of
us [Ni] to be isomorphic to to our Hopf spin chain, provided we also
require our Hopf algebra $H$ to be quasi-triangular as in [AFFS].
In this way it has been shown in [Ni] that the coaction proposed by [AFFS]
is ill-defined and should be replaced by our construction.
\footnote{There is now a revised version [AFFS(v2, May 97)], where the
  authors acknowledged our results and corrected their errors.}

\sec{The Structure of the Observable Algebra}

In this section we describe a canonical method by means of which
one
associates an observable algebra $\A$ on the 1-dimensional lattice
to any finite dimensional
$C^*$-Hopf algebra $H$. Although a good deal of our
construction works for infinite dimensional Hopf algebras as well,
we restrict the discussion here to the finite dimensional case.
If $H=\CC G$ for some finite group $G$ then our construction
reproduces the  observable algebra of the $G$-spin model of [SzV].

In Section 2.1 we provide faithful $*$-representations of the
local observable algebras $\A(I)$ associated to finite
intervals $I$ by placing a Hilbert space $\H_{even}\sim\hat H$
on each lattice site. In this way the algebras $\A(I)$ appear
as the invariant operators under a global $H$-symmetry on
$\H_{even}\o\dots\o\H_{even}$. Similarly, we may represent the
local algebras by putting Hilbert spaces $\H_{odd}\sim H$ on
each lattice link, such that $\A(I)$ is given by the invariant
operators under a global $\hat H$-symmetry on
$\H_{odd}\o\dots\o\H_{odd}$.

This is a generalization of duality transformations to Hopf spin
chains. We point out that similarly as in [SzV] both symmetries
combine to give the Drinfeld double $\D(H)$ as --- what will later
be shown to be --- the {\it universal (co-)symmetry} of our
model.

In Section 2.2 we view the Hopf spin chain in the more general
setting of algebraic quantum field theory (AQFT) as a local net.
We then introduce the notion of a {\it Quantum Gibbs system} as
a family of conditional expectations $\eta_I:\A\to\A(I)'\cap\A$
with certain consistency relations, which allow to prove that
our model satisfies a lattice version of (algebraic) Haag
duality.

\subsection{Local Observables and Order-Disorder Symmetries}

Consider $\ZZ$, the set of integers, as the set of cells of the
1-dimensional lattice: even integers represent lattice sites, the
odd ones represent links. Let $H=(H,\Delta,\varepsilon,S,*)$ be a
finite dimensional $C^*$-Hopf algebra (see Appendix A).
We denote by $\hat H$
the dual of $H$ which is then also a $C^*$-Hopf algebra. We denote
the structural maps of $\hat H$  by the same symbols
$\Delta,\varepsilon,S$.
Elements of $H$ will be
typically denoted as $a,b,\dots$, while those of $\hat H$ by
$\varphi,\psi,\dots$. The canonical pairing between $H$ and
$\hat H$
is denoted by $a\in H, \varphi\in\hat H\mapsto\langle
a,\varphi\rangle \equiv\langle\varphi,a\rangle\in\CC $.
We also identify $\hat{\hat H}=H$ and emphasize that $H$ and $\hat
H$ will always appear on an equal footing.
 There are natural left and right actions of $H$
on $\hat H$ (and vice versa) denoted by  Sweedler's arrows:
\begin{eqnarrayabc}
 a\rightarrow\varphi &=& \varphi_{(1)}\langle a,\varphi_{(2)}\rangle\\
 \varphi\leftarrow a &=& \langle
 \varphi_{(1)},a\rangle \varphi_{(2)}
\end{eqnarrayabc}
Here we have used the short cut notations $\Delta(a)=a_{(1)}\o
a_{(2)}$ and $\Delta(\varphi)=\varphi_{(1)}\o \varphi_{(2)}$
implying a summation convention in $H\o H$ and $\hat H\o\hat
H$, respectively. For a summary of definitions on Hopf algebras
and more details on our notation see Appendix A.

We associate to each even integer $2i$ a copy $\A_{2i}$ of the
$C^*$-algebra $H$ and to each odd integer $2i+1$ a copy
$\A_{2i+1}$ of $\hat H$.
We denote the elements of $\A_{2i}$ by $A_{2i}(a),\ a\in H$,
and the elements of  $\A_{2i+1}$ by $A_{2i+1}(\psi),\
\psi\in\hat H$.
The quasilocal algebra $\A_{loc}$ is defined to be the unital
*-algebra with generators $\A_{2i}(a)$ and $A_{2i+1}(\psi),
\  a\in H,\ \psi\in\hat H,\ i\in\ZZ$ and commutation relations
\begin{eqnarrayabc}
  AB&=&BA,\quad A \in \A_i,\ B\in\A_j,\  |i-j|\geq 2\\
  A_{2i+1}(\varphi)A_{2i}(a) &=& A_{2i}(a_{(1)})\langle
a_{(2)},\varphi_{(1)}\rangle A_{2i+1}(\varphi_{(2)})\\
A_{2i}(a)A_{2i-1}(\varphi) &=& A_{2i-1}(\varphi_{(1)})
\langle\varphi_{(2)},
  a_{(1)}\rangle A_{2i}(a_{(2)})
\end{eqnarrayabc}
Equation (2.2b) can be inverted to give
\eqbox{}{
 A_{2i}(a) A_{2i+1}(\varphi) &=&
 A_{2i}(a_{(3)})A_{2i+1}(\varphi)A_{2i}(S(a_{(2)})a_{(1)})\\
&=& A_{2i}(a_{(4)})A_{2i}(S(a_{(3)}))\langle S(a_{(2)}),
\varphi_{(1)}\rangle A_{2i+1}(\varphi_{(2)})A_{2i}(a_{(1)})\\
&=& \langle S(a_{(2)}),\varphi_{(1)}\rangle
A_{2i+1}(\varphi_{(2)})A_{2i}(a_{(1)})
}
and similarly for (2.2c).
Using equ. (A.3) this formula can also be used to check that the
relations (2.2b,c) respect the *-involution on $\A_{loc}$.
We denote $\A_{n,m}\subset\A_{loc}$ the unital *-subalgebra
generated by $\A_i,\ n\le i\le m$.
For $m<n $ we also put $\A_{n,m}=\CC\1$.

\smallskip
The above relations
define what can be called a two-sided iterated crossed
product, i.e.
$$\A_{n-1,m+1}=\A_{n-1} \lcros \A_{n,m}\cros\A_{m+1}$$
where $\A_{m+1}$ acts on $\A_{n,m}$ from the left via
\beq A_{m+1} (a)\; \triangleright\;
\A_{n,m} =A_{m+1} (a_{(1)})\A_{n,m}  A_{m+1} (S(a_{(2)})) \eeq
and $\A_{n-1}$ acts
on $\A_{n,m}$ from the right via
\beq\A_{n,m} \;\triangleleft\; A_{n-1}(a)=A_{n-1}
(S(a_{(1)})) \A_{n,m} A_{n-1} (a_{(2)})\eeq
and where for all $n\le m$ these two
actions commute.

We now provide a *-representation of
$\A_{n,m}$ on finite dimensional Hilbert spaces
$\H_{n,m}$ proving that the
algebras $\A_{n,m}$ are in fact finite dimensional
$C^*$-algebras and that they arise
as the invariant subalgebras in $\H_{n,m}$
under a global $H$-symmetry.
Let $h\in H$ be the
unique normalized Haar measure on $\hat H$, i.e. $h^2=h^* =h$
and $h\to \varphi =\varphi\leftarrow h =\langle h,\varphi\rangle \varepsilon$
for all $\varphi \in
\hat H$. We introduce the Hilbertspace $\H=L^2(\hat H,h)$ to be the $\CC$-
vector space $\hat H$ with scalar product
\beq\langle \varphi|\psi\rangle  := \langle h,\varphi^*\psi\rangle \eeq
Elements of $\H$ are denoted as $|\psi\rangle ,\psi \in \hat H$.
Following the notation
of [N] we introduce the following operators in $\End\H$
\eqbox{}{
Q^+(\varphi)|\psi\rangle  &:=& |\varphi\psi\rangle \\
Q^-(\varphi)|\psi\rangle  &:=& |\psi\varphi\rangle \\
P^+(a)|\psi\rangle &:=& |a\to\psi\rangle \\
P^-(a)|\psi\rangle &:=& |\psi\leftarrow a\rangle
}
where $a\in H$ and $\varphi,\psi\in\hat H$.
Using the facts that on finite dimensional $C^*$-Hopf algebras $h$ is tracial,
$S(h)=h$ and $S^2=id$ [W] one easily checks that
\eqbox{}{
Q^\pm (\varphi)^* &=& Q^\pm(\varphi^*)\\
P^\pm (a)^* &=& P^\pm(a^*)
}
Moreover $Q^\pm (\hat H)' =Q^\mp(\hat H)$ and $P^\pm (H)' =P^\mp (H)$,
where the prime denotes the commutant in
 $\End\H$. We also recall the well known
fact (see [N] for a review) that $Q^\sigma(\hat H) \vee P^{\sigma'}(H)=
\End\H$ for any choice of $\sigma,\sigma'\in \{ +,-\}$.

We now place a copy  $\H_n\simeq \H$ at each even lattice site, $n\in 2\ZZ$,
and for $n\le m$ and $n,m\in 2\ZZ$ we put
\beq\H_{n,m} := \H_n\otimes \H_{n+2} \otimes ...\otimes \H_m\eeq
We also use the obvious
notations $Q^\pm_\nu (a)$ and $P^\pm_\nu(\varphi)$ to
denote the operators acting on the tensor factor $\H_\nu,\ \nu\in 2\ZZ$.
Let now $R_{n,m}$ be the global right action of $H$ on
$\H_{n,m}$ given by
\beq R_{n,m} (a)=\prod_{i=0}^{m-n\over 2}
P^-_{n+2i} (a_{(1+i)}) ~~,~~a\in H.\eeq
and put  $L_{n,m} := R_{n,m}\circ S $ .
We then have \\
\\
{\bf Proposition 2.1:}
{\sl Let $n,m\in 2\ZZ,\ n\le m$, and let $\pi_{n,m}:
\A_{n,m} \to \End\H_{n,m}$ be given by
\eqbox{}{
\pi_{n,m} (A_{2i}(a))&=& P_{2i}^+ (a)\\
\pi_{n,m} (A_{2i+1}(\varphi)) &=& Q^-_{2i}
(S(\varphi_{(1)}) ) Q^+_{2i+2} (\varphi_{(2)})
}
Then $\pi_{n,m}$ defines a faithful *-representation of $\A_{n,m}$ on
 $\H_{n,m}$ and $\pi_{n,m}(\A_{n,m})=L_{n,m} (H)'$.}

\bigskip\noindent
{\it Proof:} We proceed by induction over $\nu={m-n\over 2}$. For $\nu=0$
the claim follows from $\pi_{n,n}(\A_{n,n})=P^+_n(H)=P^-_n (H)'$. For
$\nu\ge 1$ we use the Takesaki duality theorem for double cross products
[Ta,NaTa] saying that $\A_{n,m+2}\simeq \A_{n,m}\otimes
\End \H \simeq
\A_{n,m}\otimes \A_{m+1,m+2}$ where the isomorphism is given by
(see equ. (A.10) of Appendix A)
\eqbox{}{
\T : \A_{n,m+2}& \rightarrow& \A_{n,m} \otimes \End\H\\
\T (A)&=&A\otimes \1\\
\T (A_m(a)) &=& A_m(a_{(1)})\otimes P^-(S(a_{(2)}))\\
 \T (A_{m+1}(\psi))&=&\1 \otimes Q^+(\psi)\\
\T(A_{m+2}(a))&=&\1 \otimes P^+(a)
}
where $ A\in \A_{n,m-1},\ a\in H$ and $\psi\in\hat H$.
Hence,
by induction hypothesis
$\hat\pi_{n,m+2} := (\pi_{n,m} \otimes id)\circ \T$
defines a faithful *-representation of $\A_{n,m+2}$ and $\hat \pi_{n,m+2}
(\A_{n,m+2})=(R_{n,m}(H)\otimes \1)'$. We now identify $\H\equiv\H_{m+2}$
and construct a unitary $\hat U\in End(\H_{n,m+2})$ such that
$\pi_{n,m+2}=\Ad\hat U\circ \hat \pi_{n,m+2}$ and $R_{n,m+2}(H)=\hat U
(R_{n,m}(H)\otimes\1)\hat U^*$ which proves our claim. To this end we put
\eqbox{}{
&U:\H_m\otimes \H_{m+2} \to \H_m \otimes \H_{m+2}&\\
&U|\varphi\otimes\psi\rangle
:= |\varphi S(\psi_{(1)})\otimes \psi_{(2)}\rangle &
}
and define
$\hat U=\1_n\otimes...\otimes \1_{m-2}\otimes U$. We leave it to the
reader to check that $U$ is unitary and satisfies
\footnote{Up to a change of left-right conventions $U$ is a version of the
pentagon operator (also called Takesaki operator or multiplicative unitary),
see, e.g. [BS].}
$$U^{-1} |\varphi\o\psi\rangle =|\varphi\psi_{(1)}\o\psi_{(2)}\rangle $$
Now $\hat U$ obviously commutes with $Q^+_m(\hat H)$ and therefore with
$\pi_{n,m}(\A_{n,m-1})\otimes \1_{m+2}$, proving
$$Ad\,\hat U\circ\hat\pi_{n,m+2}|\A_{n,m-1}=\pi_{n,m+2}|\A_{n,m-1}$$
Similarly, $\hat U$ also commutes with $P^+_{m+2}(H)$, proving
$$Ad\,\hat U\circ \hat\pi_{n,m+2} |\A_{m+2}=\pi_{n,m+2}|\A_{m+2}$$
Next, we compute
\begin{eqnarray*}
UQ^+_{m+2}(\chi) |\varphi\otimes\psi\rangle  &=& |\varphi S(\psi_{(1)})
S(\chi_{(1)})\otimes\chi_{(2)}\psi_{(2)} \rangle  \\
&=& Q_m^- (S(\chi_{(1)})) Q^+_{m+2} (\chi_{(2)})U|\varphi\otimes\psi\rangle
\end{eqnarray*}
and
\begin{eqnarray*}
UP_m^+(a_{(1)}) P^-_{m+2} (S(a_{(2)}))
\lt|\rt\varphi\otimes\psi\rangle  \\
&=&\langle a_{(1)} , \varphi_{(2)}\rangle  \langle
  S(a_{(2)}),\psi_{(1)}\rangle  |\varphi_{(1)} 
S(\psi_{(2)})\otimes \psi_{(3)}\rangle \\
&=&\langle a , \varphi_{(2)}S(\psi_{(1)})\rangle  |\varphi_{(1)} S(\psi_{(2)}
\otimes \psi_{(3)}\rangle \\
&=& P_m^+(a) U|\varphi \otimes \psi\rangle
\end{eqnarray*}
proving that
$$Ad\,\hat U\circ \hat\pi_{n,m+2} | \A_{m,m+1} = \pi_{n,m+2}|\A_{m,m+1}$$
and therefore $\pi_{n,m+2} =Ad\,\hat U\circ \hat \pi_{n,m+2}$. Finally
\begin{eqnarray*}
UP_m^-(a) U^*|\varphi \otimes \psi\rangle  &=&
\langle a,\varphi_{(1)}\psi_{(1)}\rangle  U|\varphi_{(2)}
\psi_{(2)}\o \psi_{(3)}\rangle \\
&=& \langle a_{(1)} , \varphi_{(1)} \rangle \langle  a_{(2)}, \psi_{(1)}\rangle  | \varphi_{(2)}
\otimes \psi_{(2)}\rangle \\
&=& P_m^- (a_{(1)}) P_{m+2}^- (a_{(2)}) | \varphi\otimes\psi\rangle
\end{eqnarray*}
which proves $R_{n,m+2} = \Ad\hat U\circ (R_{n,m} \otimes \1_{m+2})$.
\quad {\it Q.e.d.}

\bigskip
We remark at this point that iterated application of the Takesaki duality
theorem immediately implies $\A_{i,j} \simeq (End~\H)^{\otimes\nu}$ whenever
$j=i+2\nu+1$
and therefore the important {\em split property}
of $\A$ (see subsection 2.2).
We also remark that we could equally well interchange the role of $H$ and
$\hat H$  to define faithful *-representations $\pi_{n,m}$ of $\A_{n,m}$
for $n,m\in 2\ZZ+1$, where now $\H_{2i+1} =L^2(H,\omega),\ \omega\in\hat H$
being the Haar measure on $H$. In this way $\pi_{n,m} (\A_{n,m})$ for
$n,m\in 2\ZZ+1$
would appear as the invariant algebra under a global $\hat H$-symmetry.

\bigskip

Hence, depending on how we represent them, our local observable algebras seem
to be the invariant algebras under either a global $H$-symmetry or a global
$\hat H$-symmetry. It is the purpose of this work to show that in the
thermodynamic limit both symmetries can be reconstructed from the category
of ``physical representations" of $\A$ (i.e. fulfilling an analogue of the
Doplicher-Haag-Roberts selection criterion relative to some Haag dual vacuum
representation). In a sense to be explained below $H $ and
$\hat H$ then reappear as {\it cosymmetries} of $\A$.
Generalizing and improving the methods and results of [SzV] we
will in fact prove that $H$ and $\hat H$ combine to yield the
{\it Drinfeld double}
$\D(H)$ (see Appendix B for a review of definitions) as the
{\it universal cosymmetry}
of $\A$.

\bigskip

This should be understood as a generalization of the ``order-disorder"
symmetries in G-spin
quantum  chains, which are well known to appear for finite
abelian groups $G$ and
which have been generalized to finite nonabelian groups
$G$ by [SzV]. The relation with our present formalism is obtained
by letting $H=\CC G$ be the group algebra. We then get
$\hat H=Fun (G)$,
the abelian algebra of $\CC$-valued functions on $G$, and
$\H=L^2 (G,h)$, where $h=|G|^{-1} \Sigma_g\ g\in \CC G$ is the Haar
measure on $\hat H$. Hence $\H_{n,m}\cong L^2 (G^{m-n\over
2}),\  m,n\in 2\ZZ$, and
$\pi_{n,m}$ acts on $\psi\in \H_{n,m}$ by
\begin{eqnarray*}
(\pi_{n,m}(A_{2i} (a)) \psi)(g_n,...,g_{2i},...,g_m)
&=&\psi(g_n,...,g_{2i} a,...,g_m)\\
(\pi_{n,m} (A_{2i+1}(\varphi))\psi)(g_n,...g_m)
&=& \varphi (g_{2i}^{-1}g_{2i+2})\psi(g_n,...,g_m)
\end{eqnarray*}
These operators are immediately realized to be invariant under the global
$G$-spin rotation
$$(L_{n,m} (a) \psi)(g_n,...,g_m)=\psi(a^{-1} g_n,...,a^{-1} g_m) ,~~a\in G.$$
which would then be called the ``order symmetry".

In this representation a ``disorder-symmetry" can be defined as an action
$\hat L_{n,m}$ of $\hat H = Fun (G)$
$$(\hat L_{n,m} (\varphi)\psi)(g_n,...,g_m):= \varphi (g_ng_m^{-1}) \psi
(g_n,...,g_m)$$
and it has been shown in [SzV] that $L_{n,m}$ and $\hat L_{n,m}$ together
generate a representation of the Drinfeld double $\D(G)$. Note that in the
limit $(n,m)\to (-\infty,\infty)$ all local observables are also invariant
under (i.e. commute with) $\hat L_{n,m} (\hat H)$. The generalization of
$\hat L_{n,m}$  to arbitrary finite dimensional $C^*$-Hopf algebras is given
by\\
\\
{\bf Lemma 2.2.:}
{\sl Let $n,m\in 2\ZZ,\ m\ge n+2$, and let $\hat L_{n,m} :\hat H\to End
(\H_{n,m})$ be the *-representation given by
\beq\hat L_{n,m} (\varphi) =Q_n^+ (\varphi_{(1)}) Q_m^- (S(\varphi_{(2)}))\eeq
Then $L_{n,m}(H)$ and $\hat L_{n,m} (\hat H)$ generate a
faithful *-representation
of the Drinfeld double $\D(H)$ on $\H_{n,m}$.}\\
\\
{\it Proof:}
Since $L_{n,m}$ and $\hat L_{n,m}$ define faithful *-representations
of $H$ and $\hat H$, respectively, we are left to show (see eqn. (B.1c)):
$$L_{n,m} (a_{(1)}) \langle a_{(2)},\varphi_{(1)}\rangle  \hat L_{n,m}
(\varphi_{(2)}) =
\hat L_{n,m} (\varphi_{(1)})\langle \varphi_{(2)}, a_{(1)}\rangle  L_{n,m} (a_{(2)})$$
for all $a\in H$ and $\varphi\in \hat H$. For $m=n+2$ this is a straight
forward calculation using the ``Weyl algebra relations" [N]
\begin{eqnarray*}
P^-(a)Q^+(\varphi)
&=& Q^+(\varphi_{(2)}) P^-(a_{(2)} ) \langle a_{(1)},\varphi_{(1)}\rangle \\
P^-(a)Q^-(\varphi)
&=& Q^-(\varphi_{(2)})P^-(a_{(1)} )\langle a_{(2)},\varphi_{(1)}\rangle
\end{eqnarray*}
and the identities $\Delta\circ S=(S\otimes S)\circ \Delta_{op}$
and $S^2 =id$.
For $m\ge n+4$ we proceed by induction and define the unitary
$$V:\H_{m-2}\otimes \H_m \to \H_{m-2} \otimes \H_m$$
$$V|\varphi\otimes \psi\rangle  :=|S(\psi_{(1)})\otimes \psi_{(2)}\varphi\rangle $$
Then $VQ^-_{m-2} (\varphi)=Q^-_m(\varphi)V$ and
$VP^-_{m-2} (a) = P_{m-2}^-(a_{(1)}) P^-_{m-2} (a_{(2)})V$
for all $\psi\in \hat H$ and $a\in H$.
Hence
\begin{eqnarray*}
\Ad\hat V\circ (L_{n,m-2} \otimes \1_m)&=& L_{n,m}\\
\Ad\hat V\circ (\hat L_{n,m-2} \otimes \1_m)&=& \hat L_{n,m}\\
\end{eqnarray*}
where $\hat V=\1_n\otimes\cdots\otimes \1_{m-4} \otimes V$, which proves the
claim by induction. \quad {\it Q.e.d.}

\bigskip
We remark that interchanging even and odd lattice sites in Lemma 2.2
we similarly obtain a representation of $\D(\hat H)$.
Now recall
that for abelian groups $G$ there is a well known duality transformation
which consists of
interchanging the role of $H=\CC G$ and $\hat H=\CC \hat G$ by simultaneously
also interchanging the role of even an odd lattice sites and of order and
disorder symmetries,
respectively. For nonabelian groups $G$ the dual algebra
$\hat H$ is no longer a group algebra and at first sight
the good use or even the notion of
a duality transformation seems to be lost.
It is the advantage of our more general
Hopf algebraic framework to restore this apparent asymmetry and treat both,
$H$ and $\hat H$, on a completely equal footing.
In particular we also point out that as algebras the Drinfeld doubles
$\D(H)$ and $\D(\hat \H)$ coincide (it is only the coproduct which
changes into its opposite, see Appendix B).
Hence, from an algebraic point of view
there is no intrinsic difference between
"order" and "disorder" (co-)symmetries. Distinguishing one from the other
only makes sense with respect to a particular choice of the  representations
given in Lemma 2.2 on the Hilbert spaces associated with
even or odd lattice sites, respectively.


\subsection{$\A$ as a Haag Dual Net}

\medskip
The local commutation relations (2.3) of the observables suggests
that our Hopf spin model can be viewed in the more general
setting of algebraic quantum field theory (AQFT) as a local net.
More precisely
we will use an implementation of AQFT appropriate to study lattice
models in which the local algebras are finite dimensional.
Although we borrow the language and philosophy of AQFT, the
concrete mathematical notions we need on the lattice are quite
different from the analogue notions one uses in QFT on Minkowski
space.

Let $\I$ denote the set of closed finite
subintervals of $\RR$ with endpoints in $\ZZ+{1\over 2}$.
 A net of finite dimensional $C^*$-algebras, or shortly a
{\it net} is a correspondence $I\mapsto\A(I)$ associating to each
interval $I\in\I$ a finite dimensional $C^*$-algebra $\A(I)$
together with
unital inclusions $\iota_{J,I}\colon\A(I)\to\A(J)$,
whenever $I\subset J$, such that for all $I\subset J\subset K$
one has $\iota_{K,J}\circ\iota_{J,I}=\iota_{K,I}$.
For $I=\emptyset$ we put $\A(\emptyset)=\CC\1$.

The inclusions $\iota_{J,I}$ will be suppressed and for $I\subset
J$ we will simply write $\A(I)\subset\A(J)$. If
$\Lambda$ is any (possibly infinite) subset of $\RR$
we write $\A(\Lambda)$ for the $C^*$-inductive limit of
$\A(I)$-s with $I\subset \Lambda$:
$$\A(\Lambda):=\vee_{I\subset\Lambda} \A(I).$$
Especially let $\A =\A(\RR)$. As a dense subalgebra of $\A$ we denote
$$\A_{loc} =\cup_{I\in\I} \A(I).$$
The choice of the
lattice $\ZZ+{1\over 2}$ (in place of $\ZZ$ , say) is merely
a matter of notational convenience.
In the case of our Hopf spin model we put
$$\A(I)=\vee_{i\in I\cap\ZZ}\ \A_i$$
and $\A(I)=\CC\1$ if $I\cap\ZZ=\emptyset$.

Next, for $\Lambda \subset \RR$ let
$\Lambda'= \{ x\in \RR| dist(x,\Lambda)
\ge 1\}$ which is the analogue of the ``spacelike
complement" of $\Lambda$ (for $\Lambda=\emptyset$ put
$\Lambda'=\RR$).
The net $\{\A(I)\}$ is called {\it local} if
$I\subset J'$ implies $\A(I) \subset
\A(J)',\ \forall I,J\in\I$,
where for $\B\subset\A$ we denote $\B'\equiv
\B'\cap \A$ the commutant of $\B$ in $\A$. For $\Lambda \subset \RR$ we also
denote
\eqbox{}{
\Lambda^c&:=& \RR \setminus \Lambda\\
\bar\Lambda &:=& \Lambda^{'c}\\
\Int\Lambda &:=& \Lambda^{c'} \\
\partial \Lambda
&=& \bar\Lambda \setminus \Int\Lambda = \bar\Lambda \cap \Lambda^c
}
The net $\{\A(I)\}$ is called {\it split} if for all $I\in\I$
there exists a $J\in\I$ such that $J\supset I$ and $\A(J)$ is
simple. The net is called additive, if
$\A (I)\vee\A(J) =\A(I\cup J)$ for all $I,J\subset I$, where
$M\vee N$ denotes the $C^*$-subalgebra of $\A$ generated by
the subalgebras $M,N \subset\A$.
The net is said to satisfy the intersection property if
$\A (I)\cap \A(J)=\A(I\cap J)$ for all $I,J\in\I$.

The local observable algebras $\{\A(I)\}$ of the Hopf spin model
defined in subsection 2.1 provide an example of a local additive
split net with intersection property. What is not so obvious is that this
net satisfies {\it algebraic Haag duality}.

\bigskip\noindent
{\bf Definition 2.3:} The net $\{ \A (I)\}$ is said to satisfy
(algebraic) Haag duality if
$$\A(I')' =\A (I)~~~\forall I\in \I$$

\medskip
To prove Haag duality for our model it is useful to introduce a non-commutative
analogue of a family of local Gibbs measures in
classical statistical lattice models.\\
\\
{\bf Definition 2.4:} A {\it quantum Gibbs system }
on the net $\{\A (I)\}$ is a
family of conditional expectations
$\eta_I:\A\to\A(I)'$ such that for all
$I,J\in\I$ the following conditions hold\\
$
\ba{rrcll}
i)&\qquad \eta_I\circ \eta_J&=&\eta_I,\quad &\mbox{if $J\subset I$}\\
ii)&\qquad \eta_I(\A(J))&\subset& \A(I'\cap J),\quad
&\mbox{if $I\not\subset J$}
\ea
$

\bigskip
We will now show that the existence of a quantum Gibbs system on
$\{\A(I)\}$ is already sufficient to prove Haag duality.
Since we think that
our methods might also be useful in higher dimensional models,
we will keep
our arguments quite general.
First we introduce a {\it wedge} W
as the union
$$W=\cup_n I_n$$
where $I_n\subset I_{n+1}$ is an
unbounded increasing sequence in $\I$ with the
so-called {\it wedge property}
saying that for all $J \in\I$ the sequence $I_n'\cap J$
eventually becomes constant.
Putting $W'=\cap_n I'_n$ we now have the following

\bsn
{\bf Proposition 2.5:} {\sl Assume that
the net $\{\A(I)\}$ admits a quantum Gibbs system\\
$\eta_I:\A\to \A(I)'$. Then $\A$ satisfies
\begin{itemize}
\item[i)] Wedge duality, i.e.
$\A(W)'=\A(W')~~~\mbox{for~all~wedges~} W.$
\item[ii)] The intersection property for wedge complements, i.e.
$\A(W'\cap\Lambda)=\A(W')\cap\A(\Lambda)$
for all wedges $W$ and intervals or wedges $\Lambda$.
\item[iii)] Haag duality for intervals, i.e.
$\A(I')' =\A(I)~~~\forall I\in\I.$
\end{itemize}
}

\bigskip\noindent
{\it Proof:} i) By locality we have
$\A(W')\subset \A(W)'$. Now let $I_n\subset
I_{n+1} \in\I$ and $W=\cup_n I_n$. We define
$$\eta_W:= \lim_n \;\eta_{I_n}$$
We show that the limit exists on $\A$
and defines a conditional expectation
$\eta_W:\A\to\A(W)'$. First the limit exists pointwise on $\A(J)$ for each
$J\in\I$, since there exists $n_0>0$ such that $I_{n_0}
\not\subset J$ and
$$W' \cap J= I_n' \cap J= I_{n_0}' \cap J$$
for all $n\ge n_0$.
 Hence, by Definition 2.4i), we get for all $n\ge n_0$ and
$A\in\A(J)$
$$\eta_{I_n} (A) = \eta_{In} \circ \eta_{I_{n_0}} (A) = \eta_{I_{n_0}} (A)$$
since $\eta_{I_{n_0}}(A)\in \A(I_{n_0}'\cap J)=\A(I_n'\cap J)\subset \A(I_n)'$.
Thus $\eta_{I_n}(A)$ eventually becomes constant for all $A\in \A(J)$ and all
$J\in\I$ and we get
$$\eta_W(\A(J))\subset \A(W'\cap J) ~~~\forall J\in\I$$
Hence $\eta_W$ exists on $\A_{loc}$ and is positive and
bounded by 1 since all $\eta_{I_n}$ have this property.
 Thus $\eta_W$ may be extended to all of $\A$  yielding
$$\eta_W(\A)\subset\A(W').$$
A simple
$3\varepsilon$-argument shows that the extension still satisfies
$$\eta_W(A)=\lim_n \eta_{I_n} (A)~~~\forall A\in \A.$$
Since $I_n\subset W$ we get $\A(W)'\subset \A(I_n)'$
and hence $\eta_W(A)=A$ for all $A\in\A(W)'$. This proves $\A(W)'\subset
\A(W')$ and therefore $\A(W)'=\A(W')=\eta_W(\A)$.

ii) By the above arguments we have
$$\eta_W(\A(\Lambda))\subset\A(W'\cap\Lambda)~~\mbox{ for~ all}~
\Lambda\in\I $$
and since $\eta_W$ is a conditional expectation onto
$\A(W')=\A(W)'$ we get $\eta_W(A)=A$
for all $A\in\A(W')\cap \A(\Lambda)$ implying $\A(W')\cap \A(\Lambda)
\subset \A(W'\cap\Lambda)$. The inverse inclusion again follows from locality.
Continuity of $\eta_W$ allows to push this argument from intervals $\Lambda$
to wedges $\Lambda$.

iii) Let $I\in\I$ and let $W_1$ and $W_2$ be two wedges such that $I'=
W_1 \cup W_2'$.  Then $\A(W_1) \vee \A(W_2') \subset \A(I')$ and hence
$\A(I')'\subset \A(W_1')\cap \A(W_2)=\A(W_1'\cap W_2)=\A(I)$ where we have
used wedge duality and the intersection property for wedge complements.
\qed

\bigskip
We remark that in Proposition 2.5i) we may put $W=\RR$ to
conclude that $\A$ has trivial center,
$$
\A'=\A(\RR')=\A(\emptyset)=\CC\onne\ .
$$

We now provide a quantum Gibbs system on our Hopf spin model by defining for
any $I\in\I$ and $A\in\A$
\beq
\eta_I (A) := \sum_r
{1\over n_r}
\sum_{a,b=1}^{n_r}\
e_r^{ab}Ae_r^{ba}
\eeq
where $r$ runs through the simple components $M_r\simeq Mat(n_r)$ of $\A(I)$
and $e_r^{ab}$ is a system of matrix units in $M_r$. One immediately checks
that $\eta_I:\A\to \A(I)'$ defines a conditional expectation. Moreover
$\eta_I(\A(J))\subset \A(I)'\cap \A(J\cup I)$. We now prove\\
\\
{\bf Lemma 2.6:} {\sl The family  $(\eta_I)_{I\in\I}$
provides a quantum Gibbs system
on the Hopf spin model.}\\
\\
{\it Proof:} By continuity it is enough to prove property i) of Definition 2.2
on $\A_{loc}$.
Hence let $J\subset I$ be two intervals and let $A\in\A(\Lambda),
\Lambda\in\I$, where without loss $I\cup J\subset \Lambda$. Pick a faithful
trace $tr_\Lambda$ on $\A(\Lambda)$ and define the Hilbert-Schmidt scalar
product $\bra A|B\ket
:= tr_\Lambda(A^*B),\ A,B\in\A(\Lambda)$. We clearly have
$tr_\Lambda(B\eta_I(A))=tr_\Lambda(BA)$ for all $I\subset \Lambda,
B\in \A(I)'\cap \A(\Lambda)$ and $A\in \A(\Lambda)$. Hence, for $I\subset
\Lambda$ the restriction $\eta_I | \A(\Lambda)$ is an orthogonal projection
onto $\A(\Lambda)\cap \A(I)'$ with respect to
$\bra\cdot|\cdot\ket$. Since
$J\subset I$ implies $\A(I)'\subset \A(J)'$
we conclude
$$\eta_I |\A(\Lambda) =\eta_I \circ \eta_J | \A(\Lambda)$$
To prove property ii) let $I \not\subset J$ (implying
$I\neq\emptyset$). For $\A(J)=\CC\cdot \onne$ or $\A(I)=\CC\cdot\onne$ the
statement is trivial, hence assume  $|I|\ge1$ and
$\A(J)=\A_{i,j}$ for some $i\le j\in\ZZ$.
Using property i) the claim ii) is now equivalent to
\eqbox{}{
\eta_{i-1} (\A_{i,j}) &=& \A_{i+1,j}\\
\eta_{j+1} (\A_{i,j}) &=& \A_{i,j-1}
}
where for $I=[i-{1\over 2}, i+{1\over 2}] $ we write $\eta_I\equiv \eta_i$.
Using additivity we have $\A_{i,j} =\A_i \vee \A_{i+1,j} = \A_{i,j-1}\vee
\A_j$ and hence (2.17) is equivalent to
\beq\eta_i(\A_{i\pm1}) = \CC \cdot\1,~~~~\forall i\in \ZZ\eeq
Let us prove (2.18) for $i=$even. (For odd $i$-s the proof is
quite analogous.) Choose $C^*$-matrix units $e_r^{ab}$ of
the algebra $H$.
For $r=\varepsilon$, the trivial representation (counit) of $H$,
we have $ae_\varepsilon=e_\varepsilon a
=\varepsilon(a)e_\varepsilon$, hence $e_\varepsilon\equiv h$ is just the
integral in $H$ (see Appendix A).
We now use the following

\bsn{\bf Lemma 2.7:}
{\sl Let $\BB:=(\id\o S)(\Delta(h))\in H\o H$.
Then for finite dimensional $C^*$-Hopf algebras $H$ we have}
\beq
\BB=(S\o \id)(\Delta(h))
=\sum_r{1\over n_r}\sum_{a,b}\ e_r^{ab}\otimes e_r^{ba}
\eeq

\bsn{\it Proof:}
By the Appendix A2 of [W] the Haar measure $\omega\in\hat H$
is given on $H$ by
\beq
  \omega(e_r^{ab})=\delta^{ab}
\eeq
where the normalization is fixed to $\omega(h)=1$.
Also, $\omega\circ S=\omega$.
Let $F_\omega : H \to \hat H$ denote the Fourier transformation
\beq
\bra F_\omega(a), b\ket := \omega(ab)\equiv\omega(ba)
\eeq
Then $F_\omega= \hat S\circ F_\omega\circ S$.
The inverse  Fourier transformation is given by
\beq
F_\omega^{-1}(\psi)=(\psi\o\id)(\BB)
\eeq
(see [N] for a review on Fourier transformations)
implying $(S\o S)(\BB)=\BB$.
Let $D_r^{ab}\in\hat H$ be the basis dual to $\{e_r^{ab}\}$.
Then by (2.20)
\beq
D_r^{ab}=F_\omega(\frac{1}{n_r} e_r^{ab})
\eeq
and Lemma 2.7 follows from (2.22/23) and the identity $S^2=\id$ [W].
\qed

\bigskip
From equ. (2.19) one recognizes that $\eta_i$ evaluated on $\A_{i\pm
1}$ is nothing but
the adjoint action of the integral $h$ on the dual Hopf algebra
$\hat H$. Consider the case of $\A_{i-1}$:
\beanon
\eta_i(A_{i-1}(\varphi))&=&\sum_r{1\over n_r}\sum_{a,b}
\ A_i(e_r^{ab})A_{i-1}(\varphi)A_i(e_r^{ba})\\
&=&A_i(h_{(1)})A_{i-1}(\varphi)A_i(S(h_{(2)}))\\
&=&A_{i-1}(h\to\varphi)=\one\langle\varphi|h\rangle
\eeanon
The case of $\A_{i+1}$ can be handled similarly.
\qed

\bigskip
Summarizing: The local net $\{\A(I)\}$ of the Hopf spin model is
an additive split net satisfying Haag duality and
wedge duality. Furthermore the global observable algebra $\A$
is simple, because the split property implies that $\A$
is an UHF algebra and every UHF algebra is simple [Mu].

\medskip
We finally remark without proof that the inclusion tower
$\A_{i,j}\subset\A_{i,j+1},\ j\ge i$
(or $\A_{i-1,j} \supset
\A_{i,j},\ i\le j)$ together with the family of conditional
expectation $\eta_{j+1}:\A_{i,j} \to \A_{i,j-1}\ (\eta_{i-1}:\A_{i,j} \to
\A_{i+1,j})$ precisely arises by the basic Jones construction [J] from the
conditonal expectations $\eta_{i\pm 1} :\A_i\to \CC\cdot \1$.
In particular, putting $e_{2i}=A_{2i} (h)$ and $e_{2i+1} =A_{2i+1} (\omega)$,
where $h=h^*=h^2\in H$ and
$\omega=\omega^*=\omega^2\in\hat H$ are the normalized integrals, we find
the Temperley-Lieb-Jones algebra
\eqbox{}{
&&e_i^2=e_i^* =e_i\\
&& e_ie_j = e_j e_i,~~~|i-j|\ge 2\\
&& e_i e_{i\pm1} e_i= (\dim H)^{-1}\, e_i
}

\sec{Amplimorphisms and Cosymmetries}

In this Section we pick up the methods of [SzV] to reformulate the
DHR-theory of superselection sectors for locally finite dimensional
quantum chains using the category of amplimorphisms $\Amp\A$.

In Section 3.1 we shortly review the notions and results of [SzV] and
introduce the important concept of {\it compressibility} saying
that up to equivalence all amplimorphisms can be localized in a
common finite interval $I$.
In Section 3.2 we consider the special class of amplimorphisms given
by localized coactions of some Hopf algebra $\G$ on $\A$.
We call such coactions {\it cosymmetries}.

Sections 3.3 and 3.4 investigate some general conditions under which
universal cosymmetries exist on a given net $\A$.
Here an amplimorphism $\r$ is called {\it universal}, if
 it is a sum of pairwise inequivalent and irreducible
amplimorphisms, one from each equivalence class in $\Amp\A$.
In Section 3.3 we look at properties of {\em effective
cosymmetries} and use these to show that
a universal amplimorphism becomes a cosymmetry (with respect to
suitable coproduct on $\G$) if and only if the intertwiner space
$(\r\x\r|\r)$ is ``scalar'', i.e. contained in $\onne_\A\o\Hom(V_\r,
V_\r\o V_\r)$. With this result we can prove in Section 3.4
that universal cosymmetries always exist in models which are
{\it completely compressible}.
We show that Haag dual split nets (like the Hopf spin chain) are
completely compressible iff they are compressible.
Compressibility of the Hopf spin chain will then be stated in Theorem
3.12. It will be proven later in Section 4.2, where we show that all
amplimorphisms of this model are in fact compressible into any
interval of length two.

In Section 3.5 we investigate the question of {\em uniqueness} of universal
cosymmetries. We prove that (up to automorphisms of $\G$) universal
coactions are always {\it cocycle equivalent} where we use a more
general definition of this terminology as compared to the mathematics
literature (e.g. [Ta,NaTa]). In particular this means that the coproduct
of a universal cosymmetry $\G$ on $\A$ is only determined up to cocycle
equivalence.

In Section 3.6 we discuss two notions of translation covariance
for universal coactions and relate these to the existence of a
{\em coherently translation covariant} structure in $\Amp\A$.

\subsection {The categories $\Amp\A$ and $\Rep\A$}

In this subsection $\{\A(I)\}$ denotes a split net of
finite dimensional $C^*$-algebras which satisfies algebraic Haag
duality. Furthermore we assume that the net is {\it
translation covariant}. That is the net is equipped with a
*-automorphism $\alpha\in\Aut\A$ such that
$$\alpha(\A(I))=\A(I+2)\qquad
I\in\I\,.\eqno(3.1)$$
At first we recall some notions introduced in [SzV].
An {\it amplimorphism} of $\A$ is an injective $C^*$-algebra map
$$\mu\colon\A\to\A\otimes\hbox{End}V\eqno(3.2)$$
where $V$ is some finite dimensional Hilbert space.
If $\mu(\one)=\one\otimes 1_V$  then $\mu$ is called {\it unital}.
Here we will restrict ourselves to unital amplimorphisms since the
localized amplimorphisms in a split net are all
equivalent to unital ones (see Thm. 4.13 in [SzV]). An
amplimorphism $\mu$ is called {\it localized} within $I\in\I$ if
$$
\mu(A)=A\otimes 1_V\qquad A\in\A(I^c)
$$
where $I^c:=\RR\setminus I$.
For simplicity, from now on by an amplimorphism we will always mean a
localized unital amplimorphism.

The space of {\it intertwiners} from
$\nu\colon\A\to\A\otimes\E W$ to $\mu\colon\A\to\A\otimes\E V$
is
$$
(\mu|\nu):=\{\,T\in\A\otimes\hbox{Hom}(W,V)\,|\,\mu(A)T=T\nu(A),
\ A\in\A\,\}\eqno(3.3)
$$
Two amplimorphisms
$\mu$ and $\nu$ are called {\it equivalent}, $\mu\sim\nu$, if
there
exists an isomorphism $U\in(\mu|\nu)$, that is an intertwiner $U$
satisfying $U^*U=\one\otimes 1_W$ and $UU^*=\one\otimes 1_V$.
Let $\mu$ be localized within $I$. Then $\mu$ is called {\it
transportable}
if for all integer $a$ there exists a $\nu$ localized within
$I+2a$ and such that $\nu\sim\mu$. $\mu$ is called
{\it translation covariant} if
$(\alpha^a\otimes\id_V)\circ\mu\circ\alpha^{-a}\ \sim\ \mu$ for
all $a\in \Z$.
Clearly, translation covariance implies transportability.

Let $\Amp\A$ denote the category with objects given by the localized
unital amplimorphisms $\mu$ and with arrows from
$\nu$ to $\mu$ given by the intertwiners $T\in(\mu|\nu)$.
This category has the following {\it monoidal product} :
\setc{3}
\bea
(\mu,\ \nu)\ \mapsto\ \mu\times\nu
&:=& (\mu\otimes\id_{\End W})
\circ\nu\ \colon\ \A\to\A\otimes \E V\otimes\E W\nonumber\\
T_1\in(\mu_1|\nu_1), T_2\in(\mu_2|\nu_2) &\mapsto &
T_1\times T_2:=(T_1\otimes
1_{V_2})(\nu_1\otimes\id_{\Hom(W_2,V_2)})(T_2)\\
 &\in &\ (\mu_1\times\mu_2|\nu_1\times\nu_2)\nonumber
\eea

with the monoidal unit being the trivial amplimorphism
$\id_{\A}$. The monoidal product $\times$ is a bifunctor
therefore we have $(T_1\times T_2)(S_1\times
S_2)=T_1S_1\times T_2S_2$, for all intertwiners for which the
products are defined, and $1_{\mu}\times 1_{\nu}=1_{\mu\times\nu}$
where $1_{\mu}:=\one\otimes\id_V$ is the unit arrow at the object
$\mu:\A\to\A\otimes\E V$.

$\Amp\A$ contains {\it direct sums} $\mu\oplus\nu$ of any two
objects: $(\mu\oplus\nu)(A):=\mu(A)\oplus\nu(A)$ defines a direct
sum for any orthogonal direct sum $V\oplus W$.

$\Amp\A$ has {\it subobjects}: If $P\in(\mu|\mu)$ is a Hermitean
projection then there exists an object $\nu$ and an injection
$S\in(\mu|\nu)$ such that $SS^*=P$ and $S^*S=1_{\nu}$.
The existence of subobjects is a trivial statement in the category
of all, possibly non-unital, amplimorphisms because $\nu$ can be
chosen to be $\nu (A)=P\mu(A)$ in that case. In the category $\Amp\A$ this is a
non-trivial theorem which can be proven [SzV] provided the net is
split. An amplimorphism $\mu$ is called {\it irreducible} if the
only (non-zero) subobject of $\mu$ is $\mu$. Equivalently, $\mu$
is irreducible if $(\mu|\mu)=\CC1_{\mu}$. Since the selfintertwiner
space $(\mu|\mu)$ of any localized amplimorphism is finite
dimensional (use Haag duality to show that any $T\in(\mu|\mu)$
belongs to $\A(\Int I)\otimes\E V$ where $I$ is the interval where
$\mu$ is localized, see also Lemma 3.8 below),
the category $\Amp\A$ is {\it fully
reducible}. That is any object is a finite direct sum of
irreducible objects.
The category $\Amp\A$ is called {\it rigid} if for any object $\mu$
there exists an object $\overline\mu$ and intertwiners
$C_\mu\in(\overline\mu\x\mu\,|\,\idA)\ ,\ \overline
C_\mu\in(\mu\x\overline\mu\,|\,\idA)$ satisfying
\beq
\ba{rcl}
(\overline C_\mu^*\x\1_\mu)(\1_\mu\x C_\mu) &=&\1_\mu\\
(\1_{\overline\mu}\x \overline C_\mu^*)(C_\mu\x\1_{\overline\mu})
&=&\1_{\overline\mu}
\ea
\eeq
Two full subcategories $\Amp_1\A$ and $\Amp_2\A$ of $\Amp\A$ are
called {\it equivalent}, $\Amp_1\A\sim\Amp_2\A$, if any object
in $\Amp_1\A$ is equivalent to an object in $\Amp_2\A$ and vice
versa.
For $I\in\I$ we denote $\Amp(\A,I)\subset\Amp\A$
the full subcategory of amplimorphisms localized in $I$.
We say that $\Amp\A$ is {\it compressible} (into $I$) if there exists
$I\in\I$ such that $\Amp\A\sim\Amp(\A,I)$.
Clearly, if $\Amp\A$ is compressible into $I$ then it is
compressible into $I+2a,\ \forall a\in\ZZ$. This follows, since
the translation automorphism $\alpha\in\Aut\A$ induces an
autofunctor $\underline\alpha$ on $\Amp\A$ given on objects by
$\r\mapsto\r^\alpha:=(\alpha\o\id)\circ\r\circ\alpha^{-1}$ and
on intertwiners by $T\mapsto(\alpha\o\id)(T)$. Hence
$\underline\alpha(\Amp(\A,I))=\Amp(\A,I+2)$. Moreover, we have

\bsn
{\bf Lemma 3.1:}
{\sl Let $\Amp\A$ be compressible into $I\in\I$ and let
$J\supset I+2a$ for some $a\in\ZZ$. Then all amplimorphisms in
$\Amp(\A,J)$ are transportable.}

\bsn
{\it Proof:}
Let $\{\r_r:\A\to\A\o\End V_r\}$
be a complete list of pairwise inequivalent irreducible
amplimorhisms in $\Amp(\A,I)$ and put $\r=\oplus_r \r_r$. 
\footnote{If $\A(I)$ is finite dimensional, this sum is finite.}
Then
$\r:\A\to\A\o\G,\ \G:=\oplus_r\End V_r$, is {\it universal} in
$\Amp\A$, i.e. every $\mu\in\Amp\A$ is equivalent to
$(\idA\o\beta)\circ\r$ for some $\beta\in\Rep\G$.
Moreover, $\r^\alpha\in\Amp(\A,I+2)$ is also universal
and therefore
$\r^\alpha=\Ad W\circ(\id\o\sigma)\circ\r$ for some unitary
$W\in\A\o\G$ and some $\sigma\in\Aut\G$.
Let now $J\supset I$ and $\mu=\Ad
U\circ(\idA\o\beta)\circ\r\in\Amp(\A,J)$.
Then, by Haag duality, $U\in\A(\Int J)\o\End V_\mu$,
since $U$ must commute with $\A(J^c)\o\1$.
With $\sigma\in\Aut\G$ defined as above put $\tilde\mu:=\Ad
U\circ(\idA\o\tilde\beta)\circ\r\in\Amp(\A,J)$, where
$\tilde\beta:=\beta\circ\sigma^{-1}$.
Then
$\tilde\mu^\alpha
\equiv(\alpha\o\id)\circ\tilde\mu\circ\alpha^{-1}\in\Amp(\A,J+2)$
satifies
$$
\tilde\mu^\alpha=\Ad \tilde U\circ(\idA\o\beta)\circ\r=\Ad
(\tilde UU^*)\circ\mu,
$$
where $\tilde U=(\alpha\o\id)(U)(\idA\o\tilde\beta)(W)\in\A\o\End
V_\mu$ is unitary. Thus $\mu$ is transportable into $J+2$ and
analogously into $J-2$ and therefore into $J+2a,\ a\in 2\ZZ$.
\qed

\bigskip
We remark that even if $\mu$ was localized in $J_0\subset I$, its
transported version may in general only be expected to
be smeared over all of $I+2a$.

\medskip
Next, we recall that
the full subcategory $\Amp^{tr}\A$ of transportable amplimorphisms
is a {\it braided category}. The braiding structure is
provided by the {\it statistics operators}
$$\epsilon(\mu,\nu)\ \in\ (\nu\times\mu|\mu\times\nu)\eqno(3.6)$$
defined by
$$\epsilon(\mu,\nu):=(U^*\o\one)(\onne\o
P)(\mu\o\id)(U)\eqno(3.7)$$
where $P:\End V_\mu\o\End V_\nu\to\End V_\nu\o\End V_\mu$
denotes the permutation and
where $U$ is any isomorphism from $\nu$ to some $\tilde\nu$ such
that the localization region of $\tilde\nu$ lies to the left from
that of $\mu$. The statistics operator satisfies

\eqa{7em}{
 &\mbox{naturality:}&
 \quad\epsilon(\mu_1,\mu_2)\ (T_1\times T_2)\ =\
 (T_2\times T_1)\ \epsilon(\nu_1,\nu_2) &(3.8a)\\
&\mbox{pentagons:} &
\left\{
  \begin{array}{lll}
   \epsilon(\lambda\times\mu,\nu) &=&
  (\epsilon(\lambda,\nu)\times 1_{\mu})(1_{\lambda}\times
  \epsilon(\mu,\nu))  \\
  \epsilon(\lambda,\mu\times\nu) &=&
  (1_{\mu}\times\epsilon(\lambda,\nu))(\epsilon(\lambda,\mu)\times
  1_{\nu})
 \end{array}
\right.&(3.8b)
}

The relevance of the category $\Amp\A$ to the representation
theory of the observable algebra $\A$ can be summarized in the
following theorem taken over from [SzV].

\bigskip\noindent
{\bf Theorem 3.1. } {\sl Let $\pi_0$ be a faithful irreducible
representation of $\A$ on a Hilbert space $\H_0$ that satisfies Haag
duality (here the second prime denotes the commutant in
$\L(\H_0)$):
$$\pi_0(\A(I'))'=\pi_0(\A(I))\qquad I\in\I\,.\eqno(3.9)$$
and let $\Rep\A$ be the category of
representations $\pi$ of $\OA$ that satisfy the following
selection criterion (analogue of the DHR-criterion):
$$\exists I\in\I,\
n\in\N\,:\qquad\pi\vert_{\A(I')}\simeq
n\cdot\pi_0\vert_{\A(I')}\eqno(3.10)$$
where $\simeq$ denotes unitary equivalence. Then $\Rep\A$ is
isomorphic to $\Amp\A$.
If we add the condition that $\pi_0$ is $\alpha$-covariant
and denote by $\Rep^{\alpha}\A$ the full subcategory in $\Rep\A$
of $\alpha$-covariant representations then $\Rep^{\alpha}\A$ is
isomorphic to the category $\Amp^{\alpha}\A$ of $\alpha$-covariant
amplimorphisms.}

\bigskip

In general $\Amp^{\alpha}\A\subset\Amp^{tr}\A\subset\Amp\A$.
In the Hopf spin model we shall see in Section 4
that $\Amp^{\alpha}\A=\Amp\A$
 and that $\Amp\A$ is equivalent to $\Rep\D(H)$.

\subsection{Localized Cosymmetries}

\bigskip

For simplicity we assume from now on that $\Amp\A$ contains
only finitely many equivalence classes of irreducible objects.
For the Hopf spin model this will follow from compressibility,
see Theorem 3.12 in
Section 3.4.
Let $\{\mu_r\}$ be a list of irreducible amplimorphisms in $\Amp\A$
containing exactly one from each equivalence class .
Then an object $\rho$ is called {\it universal} if it is
equivalent to $\oplus_r\mu_r$. Define the $C^*$-algebra $\G$ by
$$\G:=\oplus_r\ \E V_r$$
then every universal object is a unital $C^*$-algebra morphism
$\rho\colon \OA\to \OA\otimes\G$.
We denote by $e_r$ the minimal central projections in $\G$.
There is a distinguished 1-dimensional block $r=\varepsilon$,
i.e. $\End V_\varepsilon\cong\CC$ associated with
the identity morphism $\idA\equiv\r_\varepsilon$ as a subobject
of $\r$.
We also denote $\varepsilon\colon\G\to\CC$ the associated
1-dimensional representation of $\G$.
Note that by construction $\G$ is uniquley determined up to
isomorphisms leaving $e_\varepsilon$ invariant. We also remark
that if $\varepsilon$ is the counit with respect to some
coproduct $\Delta\colon\G\to\G\o\G$ then $e_\varepsilon$ is
the two-sided {\it integral} in $\G$, since $xe_\varepsilon
=e_\varepsilon x=\varepsilon (x) e_\varepsilon$ for all
$x\in\G$.

Universality of $\rho$ implies that any amplimorphism $\mu$ is
equivalent to $(id\o\beta_\mu)\circ\rho$ for some representation
$\beta_\mu$ of $\G$.
In particular, there must exist a $*$-algebra morphism $\Delta_\rho
:\G\to\G\o\G$ such that $\rho\times\rho$ is equivalent to
$(\id\o\Delta_\rho)\circ \rho$
\footnote{This argument fails in locally infinite theories where one
  may have $\A(I)\cong\A(I)\o\hbox{Mat}\,(n),\ \forall n\in\NN$, in
  which case the dimensions dim$\,V_\mu$ are not an invariant of the
  equivalence classes $[\mu]$.}.
As a characteristic feature of a
Hopf algebra symmetry
we now investigate the question
whether there exists an appropriate choice of $\rho$
such that $\rho\times\rho=(\idA\o\Delta)\circ\rho$ for some {\it coassociative
coproduct} $\Delta\colon\G\to\G\o\G$. If $\rho$ can be chosen in
such a way then we arrive to the very useful notion
of a comodule algebra action.

\bigskip\noindent

{\bf Definition 3.2}: Let $\G$ be a $C^*$-bialgebra with
coproduct
$\Delta$ and counit $\varepsilon$. A {\it localized comodule algebra action}
of $\G$ on $\A$ is a localized amplimorphism $\rho\colon\A\to\A\o\G$ that is also
a coaction on $\A$ with respect to the coalgebra
$(\G,\Delta,\varepsilon)$. In other words: $\rho$ is a linear map
satisfying the axioms:
\setc{10}
\begin{eqnarrayabc}
\rho(A)\rho(B)&=&\rho(AB)\\
\rho(\onne)&=&\onne\o\one\\
\rho(A^*)&=&\rho(A)^*\\
\rho\times\rho\equiv(\rho\otimes\id)\circ\rho\
&=&\ (\id\o\Delta)\circ\rho\\
(\idA\o\varepsilon)\circ\rho&=&\idA\\
\exists I\in\I\ :\ \rho(A)&=&A\o\one\quad A\in\A(I^c)
\end{eqnarrayabc}
The coaction $\rho$ is said to be {\it universal} if it is --- as an
amplimorphism --- a universal object of $\Amp\A$.

\bigskip
\noindent
For brevity by a coaction we will from now on mean a
 localized comodule algebra action in the sense of Definition 3.2.
If $\A$ admits a coaction of $(\G,\varepsilon,\Delta)$ then we also
call $\G$ a {\em localized cosymmetry} of $\A$.
Examples of universal localized cosymmetries for the Hopf spin
chain will be given in Section 4.

Next, we recall that every coaction $\rho\colon\A\to\A\o\G$ uniquely
determines
an action of the dual $\hat\G$ on $\A$, also denoted by $\rho$, as
follows (for simplicity assume $\G$ to be finite dimensional ):
\eqbox{}{
\rho_{\xi}&\colon&\A\to\A\qquad\xi\in\hat\G\\
\rho_{\xi}(A)&:=&(\idA\o \xi)(\rho(A))
}
The following axioms for a localized action of the
bialgebra $\hat\G$ on the $C^*$-algebra $\A$ are easily verified
\begin{eqnarrayabc}
\rho_{\xi}(AB)&=&\rho_{\xi_{(1)}}(A)\rho_{\xi_{(2)}}(B)\\
\rho_{\xi}(\onne)&=&\hat\varepsilon(\xi)\onne\\
\rho_{\xi}(A)^*&=&\rho_{\xi_*}(A^*)\\
\rho_{\xi}\circ\rho_{\eta}&=&\rho_{\xi\eta}\\
\rho_{\varepsilon}&=&\idA\\
\exists I\in\I\ :\ \rho_{\xi}(A)&=&\hat\varepsilon(\xi)A\,,
\quad\forall A\in\A(I^c)
\end{eqnarrayabc}
Here $\hat\varepsilon\equiv\1\in\G$ denotes the counit on
$\hat\G$.
Converseley, if $\rho_\xi$ satisfies (3.13) then
$$A\mapsto\rho(A)=\sum_s\ \rho_{\eta_s}(A)\o Y^s\ \in\ \A\o\G$$
defines a coaction, where $\{\eta_s\}$ and $\{Y^s\}$
denote a pair of dual bases of $\hat\G$ and $\G$, respectively.
In (3.13c) we used the notation $\xi\mapsto\xi_*$ for the
antilinear involutive algebra automorphism defined by
$\langle\xi_*|a\rangle=\overline{\langle\xi|a^*\rangle}$.
If $\G$ (and therefore also $\hat\G$) has an antipode $S$, then
 $\xi^*:=S(\xi_*)\equiv S^{-1}(\xi)_*$
defines a $*$-structure on $\hat\G$.

One can also check that for
${\langle\xi|a\rangle}:=D_r^{kl}(a)$, the
representation matrix of the unitary irrep $r$  of $\G$, the matrix
$\rho_{\xi}(A)$ determines an
ordinary matrix amplimorphism $\rho_r\colon\A\to\A\o M_{n_r}$.
Whether such a $\rho_r$ is irreducible is not guaranteed in
general,
so we will call it a {\it component} of $\rho$.


\subsection{Effective Cosymmetries}

\bigskip
To investigate the conditions under which the components of a given coaction
are pairwise inequivalent and irreducible we introduce the following\\
\\
{\bf Definition 3.3} Let $\rho: \A \to \A\otimes \End V_\rho$
be an amplimorphism
and let $\A$ have trivial center.
A unital *-subalgebra $\G \subset \End V_\rho$
is called {\it effective} for $\rho$, if $\rho(\A)\subset \A\otimes \G$ and
$ (\rho_r|\rho_s) = \delta_{rs}\CC ({\bf 1}_\A \otimes {\bf
1}_{V_r})$,
where $r,s$ run through a complete set of pairwise inequivalent
representations of $\G$ and where $\rho_r = (id\otimes
r)\circ\rho$. A coaction $\rho :\A\to\A\otimes\G$ is called
effective, if $\G$ is effective for $\rho$ (with respect to
some  unital inclusion $\G\subset \End V_\rho$).

\bigskip
To see whether an effective $\G\subset
\End V_{\rho}$ exists for a given
amplimorphism $\rho$, we now introduce $\Amp_\rho \A$ as the full
subcategroy of $\Amp \A$ generated by objects which are equivalent to direct
sums of the irreducibles $\rho_r$ ocurring in $\rho$ as a subobject.
We also put $\Amp^\circ_\rho \A \subset \Amp_\rho\A$ as the
full subcategory
consisting of objects $\mu$, such that all intertwiners in $(\mu|\rho)$ are
``scalar", i.e.
$$ (\mu|\rho) \subset {\bf 1}_\A \otimes\Hom (V_\rho,V_\mu)$$
Note that the amplimorphism $\rho$ itself belongs to $\Amp_\rho^\circ \A$
iff $(\rho|\rho)\equiv\rho(\A)'=\1_\A\otimes\C_\rho$ for
some unital $*$-subalgebra $\C_\rho\subset \End V_\rho$, which also
implies $\A\o\C'_\r\cap\End V_\r\subset\r(\A)$.
We now have

\bsn
{\bf Proposition 3.4:} {\sl Let $\A$ have trivial center
and let $\rho:\A\to\A\otimes \End V_\rho$
be an amplimorphism. For a unital $*$-subalgebra
$\G\subset \End V_\rho$ the following conditions are equivalent:

i) $\G$ is effective for $\rho$

ii) $(\rho|\rho)={\bf 1}_\A\otimes \C_\rho$
and $\G=\C_\rho' \cap \End V_\rho$

iii) $\rho(\A)\subset \A\otimes \G$ and $\Rep(\G)\cong \Amp_\rho^\circ (\A)$,
where the isomorphism is given on objects

by $\beta \to (id\otimes\beta)\circ\rho$
and on intertwiners by $t\to \1_\A\otimes t$.}\\
\\
{\it Proof:} Denote $V_r$ the representation spaces of
a complete set of pairwise inequivalent
irreducible representations
$r$ of $\G$. Decomposing $V_\rho$ into irreducible subspaces under the action
of $\G$ we get a family of isometries
$$u_r:V_r\otimes \CC^{N_\rho^r} \to V_\rho$$
where $N_\rho^r\in \NN$ are nonvanishing multiplicities and where $u_r^*u_s=
\delta_{rs},\ \sum_r u_r u^*_r=\1_{V_\rho}$ and
$$gu_r=u_r(r(g)\otimes\1_{N_\rho^r})~~,~~\forall g\in\G.$$
Putting $u=\oplus_ru_r:\oplus_r(V_r\otimes\CC^{N_\rho^r})\to V_\rho$
we conclude
that $u$ is an isomorphism obeying
\begin{eqnarray*}
u^*\G u &=&\oplus_r(\End V_r\otimes\1_{N_\rho^r})\\
u^*(\G'\cap End\ V_\rho)u &=&\oplus_r(\1_{V_r}\otimes Mat(N_\rho^r))
\end{eqnarray*}
and
$$(\1_\A\otimes u^*)\rho(A)(\1_\A\otimes u)=\oplus_r(\rho_r(A)\otimes
\1_{N_\rho^r})~~,~~\forall A\in \A$$
We now prove the equivalence  i)$\Leftrightarrow$   ii).

\smallskip
\noindent i) $\Rightarrow$ ii):
Let $(\rho_r|\rho_s)=\delta_{rs} \CC (\1_\A\otimes \1_{V_r})$.
Then
$$(\1_\A\otimes u^*)(\rho|\rho)(\1_\A\otimes u)=\oplus_r(\1_A\otimes\1_{V_r}
\otimes Mat(N_\rho^r))$$
which proves $(\rho|\rho)=\1_\A\otimes \C_\rho$ where $\C_\rho=\G'\cap
End\ V_\rho$ and therefore $\G=\C_\rho'\cap End\ V_\rho$.

\smallskip\noindent
ii)$\Rightarrow$ i): If $\rho(\A)'\equiv (\rho|\rho)=
\1_\A\otimes \C_\rho$ then
$\rho(\A)\subset \rho(\A)'' =\A\otimes (\C_\rho'\cap End\ V_\rho)=
\A\otimes \G$. Let now $M\in\Hom(\CC^{N^s_\rho},\CC^{N^r_\rho})$
and $T\in (\rho_r|\rho_s)$
and put
$$T_M:= (\1_\A\otimes u_r)(T\otimes M)(\1_\A\otimes u_s^*)$$
Then $T_M\in (\rho|\rho)$ and therefore $T_M=\1_\A\otimes t_M$ for some
$t_M\in \C_\rho$. Now $\C_\rho =\G'\cap End\ V_\rho$ implies
$u^*_r\C_\rho u_s=\delta_{rs}(\1_{V_r} \otimes Mat (N_\rho^r))$
and therefore
$$ T\otimes M=\1_\A\otimes u_r^* t_Mu_s \in \delta_{rs}
(\1_\A\otimes \1_{V_r}\otimes Mat (N_\rho^r))$$
which finally yields $T\in \delta_{rs}\CC(\1_\A\otimes \1_{V_r})$.

\smallskip
Next we prove the equivalence i)+ii) $\Leftrightarrow$ iii)
by first noting
that the implication iii) $\Rightarrow$ i) is trivial.
We are left with

\smallskip\noindent
i)+ii) $\Rightarrow$ iii): We first show that $\mu\in \Amp_\rho^0 \A$ implies
$(\mu|\rho_r)\subset\1_\A\otimes\Hom(V_r,V_\mu)\ \forall r$. To this end
let $e\in\CC^{N_\rho^r}$ be a unit vector and define
$\onne_\A\o u_{r,e}\in(\rho|\rho_r)$
by
$$
u_{r,e}:V_r \to V_\rho,\quad v \mapsto u_r(v\otimes e)
$$
For any $T\in (\mu|\rho_r)$ we then put
$$T_e := T(\1_\A\otimes u_{r,e}^*)$$
Then $T_e\in(\mu|\rho)$ and therefore, by assumption ii),
$T_e=\1_\A\otimes t_e$ for some
$t_e\in\Hom (V_\rho,V_\mu)$. Using $u^*_{r,e} u_{r,e} =\1_{V_r}$ we
conclude $T=\1_\A \otimes t_e u_{r,e}$ and hence $(\mu|\rho_r)$ is scalar.
Now  $\mu$ being equivalent to a direct sum of $\rho_r$'s we must have a
family of isometries
$$w_r :V_r \otimes \CC^{N_\mu^r} \to V_\mu$$
where $N_\mu^r \in \NN_o$ are possibly vanishing multiplicities
and where
$w_r^* w_s=\delta_{rs}$ (if $N_\mu^s \not= 0),\Sigma_r w_r w_r^*=\1_{V_\mu}$ and
$$\mu(A)(\1_\A\otimes w_r)=(\1_\A\otimes w_r)(\rho_r(A)\otimes\1_{N_\mu^r}),
\quad A\in \A .$$
Hence we get $\mu=(id\otimes\beta_\mu)\circ \rho$,
where $\beta_\mu\in \Rep\G$
is given by
$$\beta_\mu(g) = \Sigma_r w_r (r(g)\otimes \1_{N_\mu^r})w_r^*$$
Next, to show that $\beta\in \Rep\G$ is uniquely determined by
$\mu= (id\otimes\beta) \circ \rho \in \Amp_\rho^0 (\A)$ we define
$$\G_\rho:=\{ (\omega\otimes id_\G)(\rho(\A))|\ \omega \in \hat\A \}
\subset \G$$
where $\hat\A$ is the dual of $\A$. Clearly the restriction
$\beta |\G_\rho$
is uniquely determined by $\mu$. Moreover
$$\1_\A \otimes(\G_\rho' \cap End~ V_\rho)
=(\1_\A\otimes End~V_\rho)\cap \rho(\A)'.$$
Since, by assumption ii), $\rho(\A)'\equiv (\rho|\rho)=\1_\A\otimes(\G'\otimes
End~V_\rho)$ we conclude
$$ \G_\rho' \cap End~V_\rho =\G' \cap End~V_\rho$$
and therefore the algebraic closure of $\G_\rho$ coincides with $\G$.
Hence, being an algebra homomorphism $\beta$ is uniquely determined by its
restriction $\beta | \G_\rho$ and therefore by $\mu$.

Finally we show that
$\1_\A\otimes(\beta|\gamma)=((id \otimes\beta)\circ\rho|
(id \otimes\gamma)\circ\rho)$
for all $\beta,\gamma\in \Rep~\G$, which
in particular implies $(id\otimes\beta)
\circ \rho\in \Amp_\rho^0 \A$ for all $\beta\in \Rep\ \G$ (put $\gamma=id)$.
By decomposing $\beta$ and $\gamma$ we get unitary isomorphisms
\begin{eqnarray*}
w_\beta &:&\oplus_r(V_r\otimes \C^{N_\beta^r}) \to V_\beta\\
w_\gamma &:&\oplus_r(V_r\otimes C^{N_\gamma^r}) \to V_\gamma
\end{eqnarray*}
obeying for $x=\beta,\gamma$
$$x(g)w_x=w_x\oplus_r(r(g)\otimes \1_{N_x^r})~~~~\forall g\in \G.$$
Hence
\begin{eqnarray*}
(\1_\A\otimes w_\beta^*)\cdot
((id\otimes\beta)\circ\rho \lt |\rt (id\otimes \gamma)\circ\rho)
\cdot(\1_\A \otimes w_\gamma)\\
&=&(\oplus_r N_\beta^r\rho_r|\oplus_s N_\gamma^s\rho_s)\\
&=&\oplus_r (\1_\A \otimes \1_{V_r}
\otimes\Hom (\CC^{N_\gamma^r},\CC^{N_\beta^r}))
\end{eqnarray*}
by assumption i), which proves $((id\otimes\beta)\circ\rho|(id\otimes\gamma)
\circ\rho)=\1_\A\otimes(\beta|\gamma).$  \quad {\it Q.e.d.}\\
\\
We are now in the position to give a rather complete characterization of
effective cosymmetries.

\bsn
{\bf Theorem 3.5:} {\sl
Let $\rho:\A\to \A\otimes End~V_\rho$ be an amplimorphism
and assume $\G\subset End~V_\rho$ to be effective for $\rho$
(implying the center of $\A$ to be trivial).
Let furthermore $\varepsilon:\G\to\CC$ be a distinguished one-dimensional
representation such that $\rho_\varepsilon :=(id\otimes\varepsilon)\circ\rho=
id_\A$. Then the following conditions A)-C) are equivalent

\begin{itemize}
\item[A)] $\Amp^\circ_\rho(\A)$ closes under the monoidal product
\item[B)] $\rho\times \rho\in\Amp_\rho^\circ (\A)$
\item[C)] There exists a coassociative
          coproduct $\Delta$ on $(\G,\varepsilon)$
          such that
         $(\rho,\Delta)$ provides an effective coaction of $(\G,\varepsilon)$
          on $\A$.
\end{itemize}

\noindent Moreover, under these conditions we have

\begin{itemize}
\item[i)] $\Delta$ is uniquely determined by $\rho$.
\item[ii)] $\Amp_\rho(\A)$ is rigid iff $\G$ admits an antipode.
\item[iii)] $\Amp_\rho(\A)$ is braided, iff
            there exists a quasitriangular element
            $R\in\G\otimes\G$.
\item[iv)] $\Amp_\rho(\A)\sim \Rep(\G)$ as strict monoidal,
(rigid, braided) categories.
\end{itemize}
}

\noindent {\it Proof:} The implication $A) \Rightarrow B)$ is
obvious, since $ \rho\in\Amp_\rho^\circ (\A)$ by Proposition
3.4ii).
To prove $B)\Rightarrow C)$ let $\Delta:\G\to \End (V_\rho\otimes V_\rho)$ such
that $\rho\times\rho=(id\otimes\Delta)\circ\rho$. Then $\Delta$
uniquely exists by Proposition 3.4iii).
Moreover ${\bf 1}_\A\otimes \G'\otimes
\G'\subset (\rho\times\rho|\rho\times\rho)$ which again by
Proposition 3.4iii)
implies $\G'\otimes\G'\subset\Delta(\G)'$ and therefore $\Delta (\G)\subset
\G\otimes\G$. The identity $\rho_\varepsilon =id_\A$ implies the
counit property
$(id_\G\otimes\varepsilon)\circ\Delta =(\varepsilon\otimes id_\G)\circ \Delta =
id_\G$
and the identity $\rho\times (\rho\times\rho)=(\rho\times\rho)\times\rho$
implies the coassociativity $(id_\G\otimes \Delta)\circ\Delta=
(\Delta\otimes id_\G)\circ \Delta$. Here we have again used that any
$\beta \in \Rep\G$ is uniquely determined by $(id_\A\otimes\beta)\circ \rho$.
To prove $C)\Rightarrow A)$ we note $\Amp_\rho^\circ (\A)\cong \Rep\G$ by
Proposition 3.4iii)
and recall that $\Rep\G$ becomes monoidal for any bialgebra
$(\G,\Delta,\varepsilon)$.

Next, part i) has already been pointed out above and part iv)
follows since any object in $\Amp_\rho(\A)$
is equivalent to an object in $\Amp_\rho^\circ(\A)$ and
therefore $\Amp_\rho(\A)\sim \Amp_\rho^\circ (\A)\cong \Rep\G$
by Proposition 3.4iii). By the same argument, it is enough to
prove parts ii)+iii) with $\Amp_\rho(\A)$ replaced
by $\Rep\G$. However, for $\Rep\G$ these statements become standard (see e.g.
[Maj2,U]) and we only give a short sketch of proofs here.
So if $\beta\in \Rep\G$
and $S:\G\to\G$ is the antipode then one defines the conjugate
representation $\bar\beta:=
\beta^T\circ S$, where $\beta^T$ is the transpose of $\beta$ acting on the
dual vector space $\hat V_\beta$. Since on finite dimensional
$C^*$-Hopf
algebras $\G$ the antipode is involutive, $S^2=\id_\G$ [W], the left
and right
evaluation maps  which make $\Rep\G$ rigid
are given by the natural pairings $\hat V_\beta
\otimes V_\beta\to \CC$ and $V_\beta \otimes \hat
V_\beta\to \CC$,respectively.
Conversely, let $\Rep\G$ be rigid and identify $\G=\oplus_r\End
V_r$, where $r$ labels the simple ideals --- and therefore the
(equivalence classes of) irreducible representations --- of
$\G$. For $X\in\End V_r\subset\G$ let $S(X)\in\End V_{\overline
r}$ be given by
$$
S(X)=(\1_{\overline r}\o\overline C_r^*)
(\1_{\overline r}\o X\o\1_{\overline r})(C_r\o\1_{\overline r})
$$
We now use that for $X\in\End V_r\subset\G$ the coproduct may
be written as $\Delta(X)=\sum_{p,q}\Delta_{p,q}(X)$ where
$\Delta_{p,q}(X)\in\End V_p\o\End V_q$ is given by
$$
\Delta_{p,q} (X) = \sum_{i=1}^{N^r_{pq}}~ t^r_{pq,i} ~X~t^{r~*}_{pq,i}
$$
where $t^r_{pq,i} \in (p\times q|r), i=1,..,N^r_{pq}$,
is an orthonormal basis
of intertwiners in $\Rep\G$. Choosing a basis in $V_p$ and using
the rigidity properties (3.5) it is now not difficult to verify
the defining properties of the antipode
$$
S(X_{(1)})X_{(2)}=X_{(1)}S(X_{(2)})=\varepsilon(X)\1
$$

To prove iii) let $R\in\G\otimes\G$
be quasitriangular and let $\alpha,\beta\in \Rep\G$. Then
$$\epsilon(\alpha,\beta)
:= \sigma_{\alpha,\beta}\circ (\alpha\otimes\beta)(R)$$
defines a braiding on $\Rep\G$, where $\sigma_{\alpha,\beta}:
V_\alpha\otimes
V_\beta \to V_\beta\otimes
V_\alpha$ denotes the permutation. Conversely, let
$\epsilon(\alpha,\beta)
\in (\beta\times\alpha|\alpha\times\beta)$ be a braiding
and denote
$$R_{r,r'} := \sigma_{r',r} \circ \epsilon (r,r') \in
\End V_r\otimes \End V_{r'}$$
Putting $R:= \oplus_{r,r'}R_{r,r'}$  and using the above
formula for the coproduct it is again straightforward to
check that $R$ is quasitriangular,
i.e.
\begin{eqnarray*}
(\Delta\o\id)(R)&=& R_{13} R_{23}\\
(\id\otimes \Delta)(R)&=& R_{13} R_{12},
\end{eqnarray*}
This concludes the proof of Theorem 3.5.
\qed

\bsn
{\bf Corollary 3.6:} {\sl Necessary for a localized effective coaction
$(\rho,\Delta)$ of $(\G,\varepsilon)$ on a net $\{\A(I)\}$ to be
transportable is that $\G$ be quasitriangular.}\\
\\
{\it Proof:} If $\rho$ is transportable then any irreducible component
$\rho_r$ is transportable and hence $\Amp_\rho\A$ is braided, see equs.
(3.6-8) and [SzV]. \qed


\subsection{Universal Cosymmetries and Complete Compressibility}

Theorem 3.5 implies that $\Amp\A\sim\Rep\G$ for a suitable
$C^*$-bialgebra $(\G,\varepsilon,\Delta)$, provided we can find a universal
object $\rho=\oplus_r\rho_r$ in ${\Amp}\A$, such that $\rho\times\rho \in
\Amp_\rho^0 \A$. In this case we call $\rho$ a {\it universal
coaction} on $\A$
and $\G$ a {\it universal cosymmetry} of $\A$.
In other words, a localized coaction $\rho :\A\to\A\otimes\G$ is
universal, if and only if it is effective and for any
$\mu\in\Amp\A$ there exists a representation
$\beta_{\mu}\in\Rep\G$ such that $\mu$ is equivalent to
$(id\otimes\beta_{\mu})\circ\rho$.

We note that a priorily universal
coactions need not exist on $\A$. However, if they do, then as an algebra
$\G$ is determined up to isomorphisms, i.e.
$$ G\simeq \oplus_r End~V_r$$
where $\rho_r :\A\to \A \otimes \End V_r$ are the irreducible components of
$\rho$.
Moreover, as will be shown in Section 3.5, universal coactions $\rho$ - and
hence the
coproduct $\Delta$    on $\G$ - are determined up to cocycle equivalence
provided they exist.

\smallskip
In this
subsection we investigate the question of {\it existence} of universal
coactions $\rho$ by analysing the condition $\rho\times\rho \in
\Amp^\circ_\rho\A$.
To this end we introduce the $\rho$-stable subalgebra $\A_\rho\subset\A$
\beq
\A_\rho:=\{A\in\A|\ \rho(A)=A\otimes\1\}
\eeq
If $\B\subset\A$ is a unital $*$-subalgebra, then we say that $\rho$ is
localized away from $\B$, if $\B\subset\A_\rho$, and we denote
the full subcategory
$$\Amp(\A|\B) = \{\rho\in\Amp\A |\ \B\subset \A_\rho\}$$
We note that intertwiners between amplimorphisms in $\Amp(\A|\B)$
are always in $(\B'\cap\A)\otimes \End V_\rho$. This follows from the more
general and obvious
fact that for any two amplimorphisms $\rho_i : \A\to\A\otimes
\End V_i,\ i=1,2$, we have
$$(\rho_1|\rho_2)\subset((\A_{\rho_1}\cap \A_{\rho_2})'\cap\A)\otimes
\Hom(V_2,V_1)$$
We also note that
$\Amp(\A|B)$ clearly closes under the monoidal product. Hence we
get the immediate

\bigskip\noindent
{\bf Corollary 3.7:} {\sl Assume $\B\subset \A$ and $\B'\cap
\A=\CC\cdot\1_\A$
and let $\rho\in\Amp(\A|\B)$ be universal in $\Amp(\A|\B)$. Then
$(\rho|\rho)=\1_\A\otimes \C_\rho$ and $\rho\times\rho \in \Amp_\rho^0\A$
and therefore $\rho:\A\to\A\otimes \G$ provides an effective
coaction, where
$\G=\C_\rho' \cap \End V_\rho$.
}

\bigskip
It is suggestive to call the resulting bialgebra
$\G=:Gal(\A|\B)$  the universal
cosymmetry or ``Galois coalgebra" (since the dual bialgebra
$\hat\G$ would be the analogue of a Galois group)
 associated with the irreducible inclusion
$\B\subset\A$. If under the conditions of Corollary 3.7 $\B=\A_\rho$,
then one might also call $\B\subset \A$ a Galois extension
(recall $\B\subset \A_\rho$  by
definition).

Motivated by these considerations we call $\Amp\A$ {\it
compressible}
relative to $\B$, if any object in $\Amp\A $ is equivalent to an
object in $\Amp(\A|\B)$.

\medskip
Coming back to our net of local algebras ${\A(I)}$
this fits with our previous terminology, i.e. $\Amp\A$ is compressible
(i.e. compressible into $\A(I)$ for some $I\in\I$),
iff it is compressible relative to $\A(I^c)$ for some $I\in\I$.
Also, $\rho$ is localized
in $\Lambda$ (or equivalently on $\A(\Lambda))$, iff it is
localized away from
$\B=\A(\Lambda^c)$. We say that $\rho$ is compressible into $\Lambda$,
if it is
equivalent to an amplimorphism localized in $\Lambda$. We also recall
our previous notation
$$\Amp(\A,\Lambda)\equiv\Amp(\A|\A(\Lambda^c))$$
Our strategy for constructing localized universal coactions in $\Amp\A$
will now be to find a suitable bounded region $\Lambda=\cup_n I_n,
I_n\in\I$, such that
$\Amp\A$ is compressible into $\Lambda$ and $\A(\Lambda^c)'\cap\A=
\CC\cdot \1$.
In this case we call
$\Amp\A$ {\it completely compressible}.
By Corollary 3.7 we are then only left with constructing a
universal object in $\Amp(\A,\Lambda)$. First we note

\bigskip\noindent
{\bf Lemma 3.8:} {\sl For $i=1,2$ let $\rho_i\in\Amp(\A,I),I\in\I$,
and let the net $\{\A(I)\}$
satisfy Haag duality. Then $\rho_i(\A(I))\subset\A(I)\otimes
\End V_{\rho_i}$ and
$(\rho_1|\rho_2)\subset\A(\Int I)\otimes\Hom(V_{\rho_2},V_{\rho_1})$.}

\bigskip\noindent
{\bf Proof:}
We use the general identiy $\rho(\A(I))\subset\rho(\A(I)')'$ and the
locality property $\A(I)'\supset\A(I')$ to conclude
\begin{eqnarray*}
\rho(\A(I))&\subset& \rho(\A(I'))'\\
&=&\A(I')'\otimes \End V_\rho\\
&=&\A(I)\otimes \End V_\rho,
\end{eqnarray*}
where we have
used $\A(I')\subset\A(I^c)\subset\A_\rho$ in the second line and
Haag duality in the third line.
Since $I^c=(\Int~I)'$ we have $\A((\Int I)')\subset\A_\rho$ for all $\rho\in
\Amp(\A,I)$ and therefore $\A'_{\rho_i}\subset \A(\Int I)$ by Haag duality,
from which
$(\rho_1|\rho_2)\subset\A(\Int I)\otimes \Hom(V_{\rho_2},V_{\rho_1})$
follows.

\qed

\medskip
We remark that for additive Haag dual nets Lemma 3.8 implies that $\Amp
(\A,I)$ is uniquely determined by $\Amp(\A(I),I)$, with arrows given by
the set of intertwiners localized in $\Int I$.

Next, if the Haag dual net $\{\A(I)\}$ is also split, then for any localized
amplimorhpism $\rho$ there exists $I\in\I$ such that $\A(I)$ is simple and
$\rho$ is localized in $\A(I)$. By Lemma 3.8, $\rho$ restricts to an
amplimorphism on $\A(I)$  and by simplicity of $\A(I)$ this restriction
must be inner, i.e.
$\rho(A)=U(A\otimes\1)U^{-1}$
for some unitary $U\in\A(I)\otimes \End V_\rho$ and all $A\in\A(I)$.
Hence $\rho':=\Ad U^{-1}\circ \rho$ is localized in $\partial I$ and
we have

\bsn
{\bf Corollary 3.9:}
{\sl Let $\{\A(I)\}$ be a split net satisfying Haag duality.
Then for any localized amplimorphism $\rho$ there exists $I\in \I$ such that
$\A(I)$ is simple and $\rho$ is compressible into $\partial I$.
In particular $\Amp\A$ is completely compressible if and only if it is
compressible. }

\bigskip
{\it Proof:} The second statement follows by noting that if $\A(I)$
is simple then $\A((\partial I)^c)'\cap\A=\CC\1$,
which follows more generally from

\bsn
{\bf Lemma 3.10:} {\sl Assume Haag duality and let $I\in\I$. Then
$$\A((\partial I)^c)' =\A(I)'\cap \A(I)$$}
{\it Proof:} We have $(\partial I)^c = I\cup I'$. Hence
$\A((\partial I)^c)' =\A (I)'\cap \A(I')'=\A (I)' \cap \A(I).$
\qed

\medskip
Compressibility of $\Amp\A$ for example
holds, if $\Amp\A$ contains only finitely many equivalence
classes of irreducible objects.
Since in general we do not know this let us
now look at the obvious inclusions
$\Amp (\A,I) \subset \Amp(\A,J)$ for all $I\subset J$. If $\A (I)$ is simple
then by Corollary 3.9 $\Amp(\A, I)\sim \Amp(\A , \partial I)$. Hence
we get

\bsn
{\bf Corollary 3.11:}
{\sl Under the conditions of Corollary 3.9 let $I_n \subset
I_{n+1} \in \I$
be a sequence such that $\A (I_n)$ is simple for all
$n$ and $\cup_n I_n =\RR.$
If the sequence $\Amp(\A,\partial I_n)$ becomes constant
(up to equivalence) for $n\ge n_0$ then
$\Amp \A$ is completely compressible, i.e. compressible into $\partial
I_{n_0}$.}

\bigskip
We now recall
that in the case of our Hopf Spin model the local algebras $\A (I)$
are simple for all intervals $I$ of even length, $|I|=2n,\ n\in \NN_o.$
In particular this holds for "one-point-intervals" $I=\{ i+{1\over 2}\}$,
where $|I|=0,\ \A (I)=\CC\1$ and $\A (\partial I)=\A(\bar
I)=\A_{i,i+1}$ (since $\Int I =\emptyset)$.
The following Theorem  implies that in this model the conditions of
Corollary 3.11 hold in fact for {\it any} choice of one-point-intervals
$I_{n_0}\subset I_n$.

\bsn
{\bf Theorem 3.12:}
{\sl If $\A$ is the observable algebra of
the Hopf spin model then $\Amp\A$ is compressible into any
interval of length two.}

\bigskip
Theorem 3.12 will be proven in Section 4.2. In Section 4.1
 we will completely analyse $\Amp (\A,I)$ for all
$|I|=2$ (i.e. $\A(I)=\A_{i,i+1},\ i\in\ZZ)$,  showing that
its universal cosymmetry
is given by the Drinfeld double $\G =\D(H)$.
We also construct a universal intertwiner from $\Amp(\A,I)$ to
$\Amp(\A,I-1)$ and thereby prove that $\Amp(\A,I)$ (and therefore
$\Amp\A$) is not only transportable, but even {\em coherently
  translation covariant} (see Def. 3.17 below and [DR1, Sec.8]).


\subsection{Cocycle Equivalences}

Given two amplimorphisms $\rho,\rho'\in {\bf \Amp}(\A,\Lambda)$ which are
both universal in ${\bf \Amp}(\A,\Lambda)$ we may without loss consider
both of them as maps $\A \to \A \otimes \G$, with a fixed $*$-algebra $\G =
\oplus_r End~V_r$ and a fixed 1-dimensional representation $\varepsilon:\G
\to End~V_\varepsilon =\CC$ such that $\rho_\varepsilon=id_\A$. However, even
if $\rho$ and $\rho'$
are both effective coactions, they  may lead to different
coproducts,
$\Delta$ and $\Delta'$, on $(\G,\varepsilon)$. Coactions with
$(\G,\varepsilon)$ fixed, but with varying coproduct $\Delta$,
will be denoted as a pair $(\rho,\Delta)$.
In order to compare such coactions
we first identify coactions $(\rho,\Delta)$
and $(\rho',\Delta')$ whenever $\rho'=(id\o\sigma)\circ\rho$ and
$\Delta'=(\sigma\o\sigma)\circ\Delta\circ\sigma^{-1}$ for some
*-algebra automorphism $\sigma:\G\to\G$ satisfying
$\varepsilon\circ\sigma=\varepsilon$.
In other words, given an effective
coaction $(\rho,\Delta)$ of $(\G,\varepsilon)$ on $\A$ , then
up to a transformation by $\sigma\in\Aut(\G,\varepsilon)$ any universal
amplimorphism in $\Amp_\rho(\A)$ will be considered
to be of the form
$$\rho'=Ad~U\circ \rho$$
where $U\in\A\otimes\G$ is a unitary satisfying $(id\otimes\varepsilon)(U)=
\1_\A$.
Decomposing $\rho=\oplus_r\rho_r$ and $\rho'=\oplus_r\rho_r'$
this implies
$\rho_r\simeq \rho'_r$
for all $r$, i.e. we have fixed an ordering convention among the
irreducibles $r$ of coinciding dimensions $d_r=dim~V_r$.

We now introduce the notion of cocycle equivalence
 for coactions
$(\rho,\Delta)$. First, we recall that two coproducts, $\Delta$ and
$\Delta'$, on $(\G,\varepsilon)$ are called  {\em  cocycle
equivalent}, if
$\Delta' =Ad\,u\circ \Delta$, where $u\in\G\otimes\G$ is a
unitary  {\em left $\Delta$-cocycle}, i.e. $u^*=u^{-1}$ and
\bealph
(\1\otimes u)(id\otimes \Delta)(u) &=&
(u\otimes\1)(\Delta\otimes id)(u)\\
(id\otimes\varepsilon)(u) &=& (\varepsilon\otimes id)(u)=\1
\eealph
The most familiar case is the one where $\Delta'=\Delta_{op}$, the opposite
coproduct, and where $u=R$ is quasitriangular.
We call $u$ a {\em right $\Delta$-cocycle}, if $u^{-1}$ is a left
$\Delta$-cocycle. Note that if $u$ is a left
$\Delta$-cocycle then $\Delta':=\Ad u\circ\Delta$ is a coassociative
coproduct on $(\G,\varepsilon)$.
If in this case $S$ is an antipode for $\Delta$ then
$S'=\Ad q\circ S$ is an antipode fore $\Delta'$, where
$q:=\sum_i a_iS(b_i)$ if $u=\sum_i  a_i\o b_i$.
Moreover, $v$ is a left
$\Delta'$-cocycle iff $vu$ is a left
$\Delta$-cocycle. In particular, $u^{-1}$ is a left
$\Delta'$-cocycle. Two left $\Delta$-cocycles  $u,v$ are called {\em
  cohomologous }, if
\beq
u=(x^{-1}\o x^{-1})\,v\,\Delta(x)
\eeq
for some unitary
$x\in\G$ obeying $\varepsilon(x)=1$. A left
$\Delta$-cocycle cohomologous to ${\1\o\1}$ is called a left {\em
  $\Delta$-coboundary}.
We now give the following

\bsn
{\bf Definition 3.13:}
Let $(\rho,\Delta)$ and $(\rho',\Delta')$
be two coactions of $(\G,\varepsilon)$ on $\A$. Then a
pair $(U,u)$ of unitaries $U\in\A\o\G$ and $u\in\G\o\G$ is called
a {\it cocycle equivalence} from $(\rho,\Delta)$ to
$(\rho',\Delta')$ if
\begin{eqnarrayabc}
U\rho(A)&=&\rho'(A)U\qquad A\in\A\\
u\Delta(X)&=&\Delta'(X)u\qquad X\in\G\\
U\times_\rho U&=&(\onne\o u)\cdot(\idA\o\Delta)(U)\\
(\idA\o\varepsilon)(U)&=&\onne_\A
\end{eqnarrayabc}
where we have used the notation
\beq U\times_\rho U = (U\o\1)(\r\o\idG)(U)\in\A\o\G\o\G\eeq
The pair $(U,u)$ is called a {\it coboundary equivalence} if in addition to
(a--d) $u$ is a left $\Delta$- coboundary.
 If $u=\one\o\one$, then $(\rho,\Delta)$ and
$(\rho',\Delta')$ are called {\it strictly equivalent}.

\bigskip

Note that equs. (3.17 c,d)
imply the left $\Delta$-cocycle conditions (3.15) for $u$.
We leave it to the reader to check that the above definitions
indeed provide equivalence relations which are preserved under
transformations by $\sigma\in\Aut(\G,\varepsilon)$.
We also remark, that to our knowledge in the
literature the terminology ``cocycle equivalence
for coactions" is restricted to the case $u=\1 \otimes \1$ and hence
$\Delta'=\Delta$ [Ta,NaTa]. (If in this case
$U=(V^{-1}\otimes \1)\rho(V)$
for some unitary $V\in\A$ then $U$ would be called a $\rho$-coboundary.)

We now have

\bsn
{\bf Proposition 3.14:} {\sl Let $(\rho,\Delta)$ be an effective coaction of
$\G=\oplus_r End~V_r$ on $\A$. Then up to
transformations by $\sigma\in\Aut(\G,\varepsilon)$ all universal coactions
$(\rho',\Delta')$
in $\Amp_\rho(\A)\ (\Amp_\rho^0(\A))$ are cocycle equivalent (coboundary
equivalent) to $(\rho,\Delta)$.}

\bsn
{\bf Proof:} Let $\rho'=\Ad U\circ \rho$ where $U\in \A\otimes \G$ is unitary
and satisfies $(id\otimes \varepsilon)(U)=\1_\A$. We then have two unitary
intertwiners
\begin{eqnarray*}
(id\otimes\Delta)(U) :\rho \times \rho &\to& (id\otimes\Delta)\circ \rho'\\
U\times_\rho U : \rho\times \rho &\to& \rho'\times \rho' =(id\otimes \Delta')
\circ\rho'
\end{eqnarray*}
Now $\G$ is also
effective for $\rho'$ and therfore
any intertwiner from $(id\otimes \Delta')
\circ\rho'$ to $(id\otimes \Delta)\circ\rho'$ must be a
scalar by Proposition 3.4iii
(consider $\Delta$ and $\Delta'$ as representations of $\G$ on
$\oplus_{r,s} (V_r\otimes V_s))$.
Hence there exists a unitary $u\in\G\otimes \G$ such that
$$U\times_\rho U=(\1_\A\otimes u)(id\otimes \Delta)(U)$$
Consequently $(U,u)$ provides a cocycle  for $(\rho,\Delta)$ and
$(id\otimes\Delta')\circ\rho'=(id\otimes (\Ad u\circ \Delta))\circ \rho'$.
By Theorem 3.5i) we conclude $\Delta'=\Ad u\circ\Delta$ and therefore
$(\rho',\Delta')$ is cocycle equivalent to $(\rho,\Delta)$. If in addition
$\rho'\in\Amp_\rho^0(\A)$ then $U=\1_\A\otimes x$ for some unitary $x\in\G$.
Hence $u=(x\otimes x) \Delta(x^{-1})$ is a coboundary.
\hfill {\it Q.e.d.}


\subsection{Translation Covariance}

In this section we study transformation properties of universal
coactions under the translation automorphisms
$\alpha^a:\A\to\A,\ a\in\ZZ$.

First note that if
$(\r,\Delta)$ is a localized coaction on $\A$ then
$(\r^\alpha,\Delta)$ also is a localized coaction, where
$\r^\alpha:=(\alpha\o\id)\circ\r\circ\alpha^{-1}$.

\bsn
{\bf Definition 3.15:}
A coaction $(\r,\Delta)$ is called {\em translation covariant}
if $(\r,\Delta)$ and $(\r^\alpha,\Delta)$  are cocycle
equivalent. It is called {\em strictly translation covariant}
if $(\r,\Delta)$ and $(\r^\alpha,\Delta)$  are strictly
equivalent.

\bigskip
If $(\r,\Delta)$ is a universal coaction in $\Amp\A$, then
$(\r^\alpha,\Delta)$ is also universal. By
Proposition 3.14, $(\r,\Delta)$ and $(\r^\alpha,\Delta)$ must
be cocycle equivalent up to a transformation by
$\sigma\in\Aut(\G,\varepsilon)$. Thus, $\r$ is translation
covariant iff we can choose $\sigma=\idG$. The following Lemma
shows that this property is actually inherent in $\Amp\A$, i.e.
independent of the choice of $\r$.

\bsn
{\bf Lemma 3.16:}
{\sl Let $(\r,\Delta)$ be a universal and (strictly)
translation covariant coaction on $\A$.
Then all  universal coactions in $\Amp\A$ are (strictly)
translation covariant.}

\bsn
{\it Proof:}
By the remark after Definition 3.13 (strict) translation
covariance is  preserved under
transformations by $\sigma\in\Aut(\G,\varepsilon)$.
Let now $(W,w)$ be a cocycle equivalence from $\r$ to
$\r^\alpha$ and let $(U,u)$ be
a cocycle equivalence from $\r$ to $\r'$.
Then
$((\alpha\o\idG)(U)WU^{-1},uwu^{-1})$ is a cocycle equivalence from $\r'$ to
$\r'^\alpha$.
\qed

\bigskip
In [NSz2] we will show (see also [NSz1]) that strict
translation covariance of a universal coaction $\r$ is
necessary and sufficient for the existence of a lift of the
translation automorphism $\alpha$ on $\A$ to an automorphism
$\hat\alpha$ on the field algebra
$\F_\r\supset\A$ constructed from $\r$, such that $\hat\alpha$
commutes with the global $\G$-gauge symmetry acting on $\F_\r$.
In continuum theories with a global gauge
symmetry under a compact group there is a related result [DR1,
Thm 8.4] stating that such a lift exists if and only if the
category of translation covariant localized endomorphisms of
$\A$ is {\em coherently translation covariant}.

We now show that in our formalism these conditions actually
concide, i.e. a universal coaction $(\r,\Delta)$ on $\A$ is
strictly translation covariant if and only if $\Amp\A$ is coherently
translation covariant. Here we follow [DR1, Sec.8] (see also
[DHR4, Sec.2]) and define

\bsn
{\bf Definition 3.17:}
We say that $\Amp\A$ is {\em translation covariant} if for any
amplimorphism $\mu$ on $\A$ there exists an assignment
$\ZZ\ni a\to W_\mu(a)\in\A\o\End V_\mu$ satisfying properties
i)-iv) below. If also v) holds, then $\Amp\A$ is called {\em
coherently translation covariant}:
\bea
i) &\qquad& W_\mu(a)\in (\mu^{\alpha^a}\,|\,\mu)\\
ii) &\qquad& W_\mu(a+b)=(\alpha^a\o\id)(W_\mu(b))W_\mu(a)\\
iii) &\qquad&
  W_\mu(a)^*=W_\mu(a)^{-1}=(\alpha^a\o\id)(W_\mu(-a))\\
iv) &\qquad& W_\mu(a)T=(\alpha^a\o\id)(T)W_\nu(a),\quad\forall
  T\in(\mu\,|\,\nu)\\
v) &\qquad& W_{\mu\x\nu}(a)=(W_\mu(a)\o\1_\nu)(\mu\o\id_\nu)(W_\nu(a))
\eea

\bigskip
In the language of categories (coherenent) translation covariance of
$\Amp\A$ means that the group of autofunctors $\alpha^a,\ a\in\ZZ$, on
$\Amp\A$ is naturally (and coherently) isomorphioc to the identity
functor.
 
To illuminate these axioms let $\pi_0:\A\to\L(\H_0)$ be a
faithful Haag dual ``vacuum" representation and let $\ZZ\ni
a\to U_0(a)\in\L(\H_0)$ be a unitary representation
implementing the translations $\alpha^a$, i.e.
\beq
\Ad U_0(a)\circ\pi_0 = \pi_0\circ\alpha^a\ .
\eeq
Then given $W_\mu(a)$ satisfying i)-iii) above the ``charged"
representation $\pi_\mu = (\pi_0\o\id_\mu)\circ\mu$ is also
translation covariant, i.e.
\beq
\Ad U_\mu(a)\circ\pi_\mu = \pi_\mu\circ\alpha^a\ ,
\eeq
where the representation $\ZZ\ni a\to U_\mu(a)\in\L(\H_0)\o\End V_\mu$
is given by
\beq
U_\mu(a) = (\pi_0\o\id)(W_\mu(a)^*)(U_0(a)\o\1_\mu)\quad .
\eeq
Conversely, if $U_\mu(a)$ is a representation of $\ZZ$
satisfying (3.25) then we may define $W_\mu(a)$ satisfying
i)-iii) of Definition 3.17 by
\beq
(\pi_0\o\id)(W_\mu(a)) = (U_0(a)\o\1_\mu)U_\mu(a)^*
\eeq
Note that by faithfulness and Haag duality of $\pi_0$ this is
well defined, since if $\mu$ is localized in $I\in\I$ and if
$J\in\I$ contains $I$ and $I-a$ then the r.h.s. of (3.27)
commutes with $\pi_0(\A(J'))\o\1_\mu$ and therefore is in
$\pi_0(\A(J))\o\End V_\mu$.
In this case property iv) of Definition 3.17 is equivalent to
\beq
(\pi_0\o\id)(T)U_\mu(a) = U_\nu(a)(\pi_0\o\id)(T),\quad\forall
T\in(\nu|\mu)
\eeq
and property v) is equivalent to
\beq
U_{\mu\x\nu}(a) = (\pi_\mu\o\id)(W_\nu(a)^*)(U_\mu(a)\o\1_\nu)
\eeq

\bsn
{\bf Proposition 3.18:}
{\sl Let $\r$ be a universal coaction of
$(\G,\Delta,\varepsilon)$ on $\A$. Then $\r$ is (strictly)
translation covariant if and only if $\Amp\A$ is (coherently)
translation covariant.}

\bsn
{\it Proof:}
Let $(W,w)$ be a cocycle equivalence from $(\r,\Delta)$ to
$(\r^\alpha,\Delta)$ and define $\ZZ\ni a\to W_\r(a)\in\A\o\G$
inductively by putting $W_\r(0)=\onne\o\1$ and
\beq
W_\r(a+1) = (\alpha\o\id)(W_\r(a))W\ .
\eeq
Then $(W_\r(a),w^a)$ is a cocycle equivalence from
$(\r,\Delta)$ to $(\r^{\alpha^a},\Delta),\ \forall a\in\ZZ$.
Moreover,
\bea
W_\r(a+b) &=& (\alpha^a\o\id)(W_\r(b))W_\r(a)\\
W_\r(a)^* &=& W_\r(a)^{-1} = (\alpha^a\o\id)(W_\r(-a))
\eea
as one easily verifies. For an amplimorphism $\mu\in\Amp\A$ let
now $\beta_\mu\in\Rep\G$ and let $T_\mu\in\A\o\End V_\mu$ be a
unitary such that
\beq
\mu = \Ad T_\mu\circ(\id\o\beta_\mu)\circ\r\ .
\eeq
We then define
\beq
W_\mu(a):=(\alpha^a\o\id)(T_\mu)(\id\o\beta_\mu)(W_\r(a))T_\mu^{-1}\ .
\eeq
Since $\beta_\mu$ is determined by $\mu$ up to equivalence, the
definition (3.34) of $W_\mu(a)$ is actually independent of the
particular choice of $T_\mu$ and $\beta_\mu$. Moreover,
$W_\mu(a)$ clearly intertwines $\mu$ and $\mu^{\alpha^a}$ and
equs. (3.20/21) follow from equs. (3.31/32). To prove (3.22)
let $T\in(\mu|\nu)$. Then
$$
T_\mu^{-1}TT_\nu\in\left((\idA\o\beta_\mu)\circ\r\,|
\,(\idA\o\beta_\nu)\circ\r\right)=\onne_\A\o(\beta_\mu|\beta_\nu)
$$
by the effectiveness of $\r$. Therefore
\beq
T=T_\mu(\onne\o t)T_\nu^{-1}
\eeq
for some $t\in(\beta_\mu|\beta_\nu)$, and (3.22) follows from
(3.34/35).

If $\r$ is even strictly translation covariant then
\beq
(W_\r(a)\o\1)(\r\o\id)(W_\r(a)) = (\id\o\Delta)(W_\r(a))\ .
\eeq
We show that this implies (3.23) for all objects in
$\Amp_\r^0\A$. By Proposition 3.4iii) the amplimorphisms in
$\Amp_\r^0\A$ are all of the form
$
\mu = (\idA\o\beta_\mu)\circ\r
$
for some $\beta_\mu\in\Rep\G$ uniquely determined by $\mu$.
Hence, by (3.34)
$$
W_\mu(a) = (\idA\o\beta_\mu)(W_\r(a))\ .
$$
Moreover, using the coaction property
$\r\x\r=(\idA\o\Delta)\circ\r$
we get
$
\mu\x\nu = (\idA\o\beta_{\mu\x\nu})\circ\r
$
where $\beta_{\mu\x\nu}=(\beta_\mu\o\beta_\nu)\circ\Delta$. Hence
\bea
W_{\mu\x\nu}(a) &=& (\idA\o\beta_{\mu\x\nu})(W_\r(a))\nonumber\\
&=&(\idA\o\beta_\mu\o\beta_\nu)\circ(\idA\o\Delta)(W_\r(a))\nonumber\\
&=&(W_\mu(a)\o\1_\nu)(\mu\o\id_\nu)(W_\nu(a))
\eea
where we have used (3.36).
This proves (3.32) in $\Amp_\r^0\A$. The  extension to
$\Amp\A\sim\Amp_\r^0\A$ follows straightforwardly from (3.22).

\smallskip
Conversely, let now $\Amp\A$ be translation covariant and
identify $\G$ with the direct sum of its irreducible
representations, $\G=\oplus_r\End V_r$. Then $\r=\oplus_r\r_r$
is a special amplimorphism and
$W_\r(a)=\oplus_rW_r(a)\in\A\o\G$ is an equivalence from $\r$
to $\r^{\alpha^a}$, which must be a cocycle equivalence by
Proposition 3.14. Hence $\r$ is translation covariant. If
moreover $\Amp\A$ is coherently translation covariant then by
(3.18) and (3.23)
\beq
W_{\r\x\r}(a) = W_\r(a)\x_\r W_\r(a)
\eeq
On the other hand, similarly as in the proof of Proposition
3.4iii) equ. (3.22) implies
$$
W_{(\idA\o\beta)\circ\r}(a) = (\idA\o\beta)(W_\r(a))
$$
for all $\beta\in\Rep\G$.
Putting $\beta=\Delta:\G\to\G\o\G$ this gives
\beq
W_{\r\x\r}(a)\equiv W_{(\idA\o\Delta)\circ\r}(a) =
(\idA\o\Delta)(W_\r(a))
\eeq
and by (3.38/39) $\r$ is strictly translation covariant.
\qed


\sec{The Drinfeld Double as a Universal Cosymmetry}

In this section we prove that the Drinfeld double $\D(H)$ is
a universal cosymmetry of the Hopf spin chain. To this end we
construct in Section 4.1 a family of "two-point" coactions
$\r_I:\A(I)\to\A(I)\o\D(H)$ for any interval $I\in\I$ of length
two. We then prove that $\r_I$ extends to a universal coaction
in $\Amp(\A,I)$. We also explicitely provide the cocycle
equivalences from $\r_I$ to $\r_{I-1}$ and show that $\r_I$
and $\r_{I-2}$ are strictly equivalent and therefore --- being translates
of each other --- also strictly translation covariant. Moreover,
the statistics operators $\epsilon(\r_I,\r_I)$ are given in
terms of the standard quasitriangular R-matrix in
$\D(H)\o\D(H)$.
Finally, for any left 2-cocycle $u\in\D(H)\o\D(H)$ we construct
a unitary $U\in\A\o\D(H)$ and a universal coaction
$(\r',\Delta')$ on $\A$ such that $(U,u) $ provides a cocycle
equivalence from $\r_I$ to $\r'$.
The statistics operator for $\r'$ is given in terms of the
twisted R-matrix $u^{op}Ru^*$.

In Section 4.2 we proceed with constructing ``edge" amplimorphisms
$\r_{\partial I}:\A(\partial I)\to\A\o\D(H)$ for all intervals $I$
of (nonzero) even length, which extend to
universal ampimorphisms in $\Amp(\A,\partial I)$.
We then show that these edge amplimorphisms are all equivalent
to the previous two-point amplimorphisms. By Corollary 3.11
this proves complete compressibility of the Hopf spin chain
as stated in Theorem 3.12. Thus the double $\D(H)$ is the
universal cosymmetry of our model.

\subsection{The Two-Point Amplimorphisms}

In this subsection we provide a universal and strictly translation covariant
coaction $\rho_I\in
\Amp(\A,I)$ of the Drinfeld double
$\D(H)$ on our Hopf spin chain $\A$ for any
interval $I$ of length  $|I|=2$. Anticipating the proof of Theorem 3.12
this proves that $\D(H)$ is the universal cosymmetry of $\A$.

A review of the Drinfeld $\D(H)$ double is given in Appendix B.
Here we just note that it is generated by $H$ and $\hat H_{cop}$
which are both
contained as Hopf subalgebras in $\D(H)$, where $\hat H_{cop}$ is the Hopf
algebra
$\hat H$ with opposite coproduct. We denote the generators of $\D(H)$
by $\D(a),\ a\in H$, and $\D(\varphi),\ \varphi\in\hat H$, respectively.

\bsn
{\bf Theorem 4.1:} {\sl On the Hopf spin chain define
$\rho_I:\A(I)\to\A(I)\otimes\D(H),\ |I|=2$, by
\footnote{Here we identify $I$ with
$I\cap\ZZ$. }
\bealph
\reven(A_{2i}(a)A_{2i+1}(\varphi))&:=&A_{2i}(a_{(1)})
A_{2i+1}(\varphi_{(2)})\ \o\ \D(a_{(2)})\D(\varphi_{(1)})\\
\rodd(A_{2i-1}(\varphi)A_{2i}(a))&:=&A_{2i-1}(\varphi_{(1)})
A_{2i}(a_{(2)})\ \o\ \D(\varphi_{(2)})\D(a_{(1)})
\eealph
Then:
\bit
\item[i)]
  $\rho_{i,i+1}$ provides a coaction of $\D(H)$ on $\A_{i,i+1}$
with respect to the natural coproducts
$\Del$ (if $i$ is even) or $\Delop$ (if $i$ is odd) on $\D(H)$.
\item[ii)] $\rho_{i,i+1}$ extends to a  coaction in
$\Amp(\A,I)$ which is universal in $\Amp(\A,I)$ .
\eit
}

\bsn
{\it Proof:} i) Since interchanging even and odd sites amounts to interchaning
$H$ and $\hat H$ and since $\D(\hat H)=\D(H)_{cop}$ it is enough to prove
all statements for $i$  even.
It is obvious that the restrictions $\rho_{2i,2i+1}
|\A_{2i}$  and $ \rho_{2i,2i+1}|\A_{2i+1}$ define *-algebra
homomorphisms. Hence, to prove that $\rho_{2i,2i+1}:
\A_{2i,2i+1}\to \A_{2i,2i+1}
\otimes \D(H)$ is  a well defined amplimorphism we are left to check
that the commutation relations (2.2) are respected, i.e.
$$\rho_{2i,2i+1} (A_{2i+1}(\varphi))\rho_{2i,2i+1} (A_{2i}(a))=
\rho_{2i,2i+1}\left(A_{2i} (a_{(1)})\bra a_{(2)},\varphi_{(1)}\ket
A_{2i+1}(\varphi_{(2)})\right)$$
Using eqn. (B.2) this is straightforward and is left to the reader.
Using equs. (B.3a,b) the identities $(\id_\A\otimes \varepsilon_\D)\circ
\rho_{2i,2i+1}=\idA$ and $(\rho_{2i,2i+1} \times \rho_{2i,2i+1})=
(\id\otimes\Del)
\circ \rho_{2i,2i+1}$ are nearly trivial and are also left to the reader.

ii) To show that $\rho_I$ extends to an amplimorphism in $\Amp(\A,I)$
(still denoted by $\rho_I$) we have to check that together with
the definition $\rho_I(A):=A\otimes \1_{\D(H)},\
A\in\A(I^c)$, we get a well defined *-algebra homomorphism
$\rho_I:\A\to\A\otimes\D(H)$. Clearly, this holds if and only if $\rho_{i,i+1}|
\A_{i,i+1}$ commutes with the left adjoint action of $\A_{i+2}$ and the right
adjoint action of $\A_{i-1}$, respectively, on $\A_{i,i+1}$, where these
actions are defined on $B\in\A_{i,i+1}$ by
\begin{eqnarray*}
A_{2i+2}(a)\triangleright B &:=& A_{2i+1}(a_{(1)})B A_{2i+1}(S(a_{(2)}))\\
B\triangleleft A_{2i-1}(\varphi)
&:=& A_{2i-1}(S(\varphi_{(1)})) BA_{2i-1}(\varphi_{(2)})
\end{eqnarray*}
Now $\A_{2i+2}$ commutes with
$\A_{2i}$ and $\A_{2i-1}$ commutes with $\A_{2i+1}$
and
\bealph
A_{2i+2}(a) \triangleright A_{2i+1}(\varphi) &=& A_{2i+1}(a\to\varphi)\\
A_{2i}(a)\triangleleft A_{2i-1}(\varphi) &=& A_{2i} (a\leftarrow \varphi)
\eealph
Hence $\rho_{2i,2i+1}$ commutes with these actions, since by coassociativity
\begin{eqnarray*}
A_{2i}((a\leftarrow\varphi)_{(1)})\otimes\D((a\leftarrow\varphi)_{(2)})&=&
A_{2i}(a_{(1)}\leftarrow\varphi)\otimes\D(a_{(2)})\\
A_{2i+1} ((a\to\varphi)_{(2)})\otimes\D((a\to\varphi)_{(1)}) &=&
A_{2i+1} (a\to\varphi_{(2)})\otimes\D(\varphi_{(1)})
\end{eqnarray*}
Next we identify $\D(H)=\oplus_r \End V_r \subset \End V$,
where $r$ runs through
a complete set of pairwise inequivalent irreducible representations of
$\D(H)$ and where $V:=\oplus_rV_r$. Since $|I|=2$ implies
$\A(\Int I)=\CC\cdot\onne_\A$
we conclude by Lemma 3.8
$$\rho_{2i,2i+1}(\A)'\cap (\A\otimes \End V)=\onne_\A\otimes \C$$
for some unital *-subalgebra $\C \subset \End V$. Hence, by Proposition 3.4ii,
$\D(H)$ is effective for $\rho_{2i,2i+1}$ provided $\C=\D(H)'\cap \End V$.
To show this we now compute for $a\in H$ and $\varphi\in\hat H$
\begin{eqnarray*}
\lefteqn{\left[A_{2i+1}(S(\varphi_{(2)}))A_{2i} (S(a_{(1)}))
\otimes\1_{\D(H)}\right]\cdot\rho_{2i,2i+1}
\left(A_{2i}(a_{(2)})A_{2i+1}(\varphi_{(1)})\right)}\\
&&=A_{2i+1}(S(\varphi_{(3)})) A_{2i}(S(a_{(1)})a_{(2)} ) A_{2i+1}
(\varphi_{(2)})
\otimes \D(a_{(3)})\D(\varphi_{(1)})\\
&&=\onne_\A \otimes \D(a)\D(\varphi).
\end{eqnarray*}
Hence, $\A\otimes \D(H)=(\A\otimes \1_{\D(H)})\vee \rho_{2i,2i+1}(\A)$ and
therefore
\begin{eqnarray*}
 \onne_\A \otimes (\D(H)'\cap \End V) &\equiv&
(\A\otimes \D(H))'\cap (\A\otimes \End V)\\
&=& (\A\otimes\1_{\D(H)})'\cap\rho_{2i,2i+1}(\A)'\cap
(\A\otimes \End V)\\
&=&\1_\A\otimes\C
\end{eqnarray*}
which proves that $\D(H)$ is effective for $\rho_{2i,2i+1}$.
To prove that $\rho_I$ is universal in $\Amp(\A,I)$ we now show
$\Amp(\A,I)\subset\Amp^0_{\rho_I}(\A)$.
Hence let $\mu\in\Amp(\A,I),\ I\cap \ZZ=\{ 2i,2i+1\}$. Then $\mu(\A_{2i,2i+1})
\subset \A_{2i,2i+1}\otimes\D(H)$ by Lemma 3.8 and the restriction $\mu|
\A_{2i,2i+1}$ commutes with the left adjoint action of
$\A_{2i+2}$ and the right
adjoint action of $\A_{2i-1}$, respectively, on $\A_{2i,2i+1}$. This allows to
construct a representation $\beta_\mu:\D(H)\to \End V_\mu$ such that
$\mu=(id\otimes\beta_\mu)\circ \rho_{2i,2i+1}$ and therefore,  by
Proposition 3.4iii),  $\mu\in\Amp^0_{\rho_{2i,2i+1}}(\A)$, as follows.
First we use the above commutation properties together with eqn (2.17)
to conclude
$$
\ba{rcccl}
\mu(\A_{2i})&\subset&(\A_{2i,2i+1} \cap\A'_{2i+2})\otimes \End V_\mu
&=&\A_{2i}\otimes \End V_\mu\\
\mu(\A_{2i+1})&\subset&(\A_{2i,2i+1} \cap\A'_{2i-1})\otimes \End V_\mu
&=&\A_{2i+1}\otimes \End V_\mu
\ea
$$
Now we define, for $a\in H\subset\D(H)$ and $\varphi\in\hat H\subset\D(H)$,
\bealph
\beta_\mu(\D(a)) &:=& (A_{2i}(S(a_{(1)}))\otimes\1)\,\mu(A_{2i}(a_{(2)}))\\
\beta_\mu(\D(\varphi))
&:=& \mu(A_{2i+1}(\varphi_{(1)}))\,(A_{2i+1}(S(\varphi_{(2)}))
\otimes\1)
\eealph
Using that $\mu$ commutes
with the (left or right) adjont actions of $\A_{2i-1}$
and $\A_{2i+2}$, respectivley,
it is straightforward to check that $\beta_\mu(H)\subset\A_{2i}\otimes
\End V_\mu$ commutes with $\A_{2i-1} \otimes \1$ and $\beta_\mu(\hat H)
\subset \A_{2i+1}\otimes \End V_\mu$ commutes with $\A_{2i+2}\otimes \1$.
 Hence, by eqn. (2.18), $\beta_\mu|H$ and  $\beta_\mu|\hat H$ take
 values in
$\1_\A\otimes \End V_\mu$ and therefore (identifying $\A_{2i}=H$ and
$\A_{2i+1}=\hat H$)
\begin{eqnarray*}
\beta_\mu|H&=&(\varepsilon_H\otimes id)\circ\beta_\mu|H=(\varepsilon_H\otimes id)
\circ\mu|\A_{2i}\\
\beta_\mu|\hat H&=&(\varepsilon_{\hat H}\otimes id)\circ\beta_\mu|\hat H=
(\varepsilon_{\hat H}\otimes id)\circ\mu|\A_{2i+1}
\end{eqnarray*}
where $\varepsilon_H$ and $\varepsilon_{\hat H}$ denote the counits on $H$ and
$\hat H$, respectively, and where the second identities follow from
the definition (4.3). Thus, identifying
$\onne_\A\otimes \End V_\mu=\End V_\mu$,
the maps $\beta_\mu|H$ and $\beta_\mu|\hat H$ define *-representations
of $H$ and $\hat H$, respectively, on $V_\mu$. Moreover, inverting
(4.3) we get
\bealph
\mu(A_{2i} (a))&=&A_{2i}(a_{(1)})\otimes \beta_\mu(\D(a_{(2)}))\\
\mu(A_{2i+1}(\varphi))&=&
A_{2i+1}(\varphi_{(2)})\otimes \beta_\mu(\D(\varphi_{(1)}))
\eealph
Thus $\mu=(id\otimes\beta_\mu)\circ\rho_I$, provided that $\beta_\mu$ actually
extends to a representation of all of $\D(H)$.
To see this we have to check that $\beta_\mu$
respects the commutation relations (B.1c). Recalling the identity
$A_{2i+1}(\varphi)A_{2i}(a)=A_{2i}(a_{(1)})\bra a_{(2)},\varphi_{(1)}\ket
A_{2i+1}(\varphi_{(2)})$ and the definition (4.3) we compute
\begin{eqnarray*}
&&\1_\A\,\otimes\,\beta_\mu(\D(a_{(1)}))\,\bra a_{(2)},\varphi_{(1)}\ket\,
\beta_\mu(\D(\varphi_{(2)}))\\
&&= \left(A_{2i}(S(a_{(1)}))\otimes\1\right)\,
\mu\left(A_{2i+1}(\varphi_{(1)})A_{2i}(a_{(2)})\right)\,
\left(A_{2i+1}(S(\varphi_{(2)}))\otimes\1\right)\\
&&= A_{2i}(S(a_{(1)}))\,A_{2i+1}(\varphi_{(2)})\,A_{2i}(a_{(2)})\,
A_{2i+1}(S(\varphi_{(3)}))\,\otimes\,
\beta_\mu(\D(\varphi_{(1)}))\,\beta_\mu(\D(a_{(3)}))\\
&&=\1_\A\,\otimes\,\beta_\mu(\D(\varphi_{(1)}))\,\bra\varphi_{(2)},a_{(1)}\ket\,
\beta_\mu(\D(a_{(2)}))
\end{eqnarray*}
where in the third line we have used (4.4). Hence, by (B.1c) $\beta_\mu$
extends to
a representation of $\D(H)$ and therefore $\mu\in\Amp^0_{\rho_I}(\A)$.
This proves that $\rho_I$ is universal in $\Amp(\A,I)$.
\qed

\bsn
We now show that the coactions $\rho_{i,i+1}$ are all cocycle
equivalent and strictly translation covariant.
To this end let $\{b_A\}$ be a basis in $H$ with dual basis $\{\beta^A\}$ in
$\hat H$ and define the {\em charge transporters} $T_i\in\A_i\otimes \D(H)$ by
\beq
T_i:=\left\{
\ba{l}
A_i(b_A)\otimes D(\beta^A)\qquad i=\hbox{even}\\
A_i(\beta^A)\otimes D(b_A)\qquad i=\hbox{odd}
\ea
\right.
\eeq
Also recall that the canonical quasitriangular R-matrix in $\D(H)\otimes\D(H)$
is given by
$$R=\D(b_A)\otimes \D(\beta^A)$$
We then have

\bsn
{\bf Proposition 4.2:}
{\sl The charge transporters $T_i$ are
unitary intertwiners from $\rho_{i,i+1}$ to $\rho_{i-1,i}$, i.e.
\beq
T_i\rho_{i,i+1}(A)\ =\ \rho_{i-1,i}(A)T_i\,,\qquad
A\in\A
\eeq
and they satisfy the cocycle condition
\beq
\ba{c}
T_i\times_{\r_{i,i+1}} T_i\
\ba[t]{ll}
\equiv &
(T_i\otimes\one)\cdot(\rho_{i,i+1}\otimes\id)(T_i)\ =\\
=&\left\{
\ba{ll}
(\one\otimes R)\cdot(\id\o\Del)(T_i)\quad & i=\hbox{even}\\
(\one\o R^{op})\cdot(\id\o\Delop)(T_i)\quad & i=\hbox{odd}
\ea
\right.
\ea\ea
\eeq}
%
{\it Proof}: This is a lengthy but straightforward calculation, which
we leave to the reader.
\hfill{\it Q.e.d.}

\bigskip
Iterating the identities (4.6/7) we get an infinite sequence of cocycle
equivalences
$$
\dots(\reven,\Del)
\stackrel{(T_{2i+1},R^{op})}{\longleftarrow}
(\rho_{2i+1,2i+2},\Delop)
\stackrel{(T_{2i+2},R)}{\longleftarrow}
(\rho_{2i+2,2i+3},\Del)\dots
$$
Composing two such arrows we obtain a coboundary equivalence
$(T_{2i+1}T_{2i+2}, R^{op}R)$ because $R^{op}R=(s\o s)
\Del(s^{-1})$ according to [Dr], where $s\in\D(H)$ is the central
unitary $s=S_\D(R_2)R_1=\D(S(\beta^A))\D(b_A)$.
Likewise $(T_{2i}T_{2i+1},RR^{op})$ yields a coboundary
equivalence. Therefore introducing
\beq
U_{i,i+1}:=(\one\o s^{-1})T_iT_{i+1}\quad\in\ (\rho_{i-1,i}|
 \rho_{i+1,i+2})
\eeq
 we obtain unitary charge transporters localized within $\{i,i+1\}$ that
satisfiy the {\em trivial cocycle} conditions
\beq
\ba{rcl}
U_{2i-1,2i}\ \times_{\r_{2i,2i+1}}\ U_{2i-1,2i}\ &=&\ (\idA\o
\Del)(U_{2i-1,2i})\\
U_{2i-2,2i-1}\ \times_{\r_{2i-1,2i}}\ U_{2i-2,2i-1}\ &=&\ (\idA\o
\Delop)(U_{2i-2,2i-1})
\ea
\eeq

Hence, summarizing the above results (and
 anticipating the result of Theorem 3.12) we have shown

\bsn
{\bf Corollary 4.3:} {\sl The  coactions $\rho_{i,i+1}$ are all
  strictly translation covariant and universal in $\Amp\A$.}

\bsn{\it Proof:}
Universality follows from Theorem 4.1ii) and Theorem 3.12 and
strict translation covariance (Definition 3.15) follows from (4.8/9), since
$\r_{i+1,i+2}=(\alpha\o\id)\circ\r_{i-1,i}\circ\alpha^{-1}$.
\qed

\bigskip
Proposition 4.2 also enables us to compute the statistics
operator of $\r_I$.

\bsn
{\bf Theorem 4.4:}
{\sl Let $\r_I$ be given as in Theorem 4.1 and let
$\epsilon(\r_I,\r_I)$ be the associated statistics operator
(3.7). Then
\beq
\epsilon(\r_I,\r_I)=\onne\o PR_I
\eeq
where $P:\D(H)\o\D(H)\to\D(H)\o\D(H)$ denotes the permutation
and
\beq
R_{i,i+1}=\left\{
\ba{lll}
R\quad & , & i=\hbox{even}\\
R^{op}\quad &,& i=\hbox{odd}
\ea
\right .
\eeq
Moreover, if $(U,u)$ is a cocycle equivalence from
$(\r_I,\Del^{(op)})$ to $(\r',\Delta')$ then
$\epsilon(\r',\r')=\onne\o PR'$ where $R'=u_{op}R_Iu^*$.}

\bsn
{\it Proof:}
Putting $I\cap\ZZ=\{i,i+1\}$ and using (3.7) and (4.8) we get
\bea
(\onne\o P)\epsilon(\r_I,\r_I) &=&
(U^*_{i-1,i})^{02}(\r_{i,i+1}\o\idG)(U_{i-1,i})\nonumber\\
&=& (T^*_i)^{02}(T^*_i)^{01}(T_i\x_{\r_{i,i+1}}T_i)\,,
\eea
where the superfix $01/02$ refers to the obvious inclusions of
$\A\o\D(H)$ into $\A\o\D(H)\o\D(H)$, and where the second line
follows since $s$ is central and
$(\r_{i,i+1}\o\idG)(T_{i-1})=T_{i-1}^{02}$. Now (4.10/11)
follows from (4.7) and (4.12) by using $\Delop=\Ad R\circ\Del$
and the identities
$$
(\idA\o\Del)(T_i)=\left\{
\ba{lll}
T_i^{02}T_i^{01}\quad & , & i=\hbox{even}\\
T_i^{01}T_i^{02}\quad &,& i=\hbox{odd}
\ea
\right .
$$
which follow straightforwardly from (4.5).

Let now $(U,u)$ be a cocycle equivalence from $(\r,\Delta)$ to
$(\r',\Delta')$. Then by (3.8a) and (3.17c)
\beanon
(\onne\o P)\epsilon(\r',\r') &=&
(\onne\o P)(U\x_\r U)\epsilon(\r,\r)(U\x_\r U)^*\\
&=&(\onne\o u^{op})(\idA\o\Delta^{op})(U)(\onne\o R)
(\idA\o\Delta)(U^*)(\onne\o u^*)\\
&=&\onne\o (u^{op}Ru^*)\,.
\eeanon
\qed

\bigskip
We conclude this subsection by demonstrating that for any left
2-cocycle $u\in\D(H)\o\D(H)$ there exists a coaction
$(\r',\Delta')$ which is cocycle equivalent to
$(\r_I,\Delta^{(op)})$. To this end we first note that there
exist $*$-algebra inclusions  $\Lambda_{i,i+1}:\D(H)\to\A$
given by
\beanon
\Lambda_{2i,2i+1} (\D(a)) &:=& A_{2i}(a)\\
\Lambda_{2i,2i+1} (\D(\varphi)) &:=& A_{2i-1}
(\varphi_{(2)}) A_{2i+1}(\varphi_{(1)})
\eeanon
and analogously for $\Lambda_{2i-1,2i}$.
Moreover, the following identities are straightforwardly checked
$$
\rho_I\circ \Lambda_I = (\Lambda_I\otimes \id)\circ\Del^{(op)}
$$
For a given 2-cocycle $u\in\D(H)\o\D(H)$ we now put
$\Delta'=\Ad u\circ\Del^{(op)},\ 
U=(\Lambda_I\o\id)(u)$ and $\r'=\Ad U\circ\r_I$, from which it
is not difficult to see that $(U,u)$ provides a cocycle
equivalence from $(\r_I,\Del^{(op)})$ to $(\r',\Delta')$.


\subsection {Edge Amplimorphisms and Complete Compressibility }

This subsection is devoted to the construction of universal edge
amplimorphisms and thereby to the proof of Theorem 3.12. As a
preparation we first need

\bsn{\bf Proposition 4.5:}
{\sl Let $j=i+2n+1,\ i\in\ZZ,\ n\in\NN_0$.
Then there exist *-algebra inclusions
$$\ba{lll}
L_{i,j} &:& \A_{i-1}\to\A_{i,j}\cap\A_{i+1,j}'\\
R_{i,j} &:& \A_{j+1}\to\A_{i,j}\cap\A_{i,j-1}'
\ea $$
such that for all $A_{i-1}(a)\in\A_{i-1}$ and all
$A_{j+1}(\varphi)\in\A_{j+1}$
\bea
i) && A_{i-1}(a_{(1)})L_{i,j}(S(a_{(2)}))\in
\A_{i-1,j}\cap\A'_{i,j}\\
ii) && R_{i,j}(S(\varphi_{(1)}))A_{j+1}(\varphi_{(2)})\in
\A_{i,j+1}\cap\A'_{i,j}\\
iii)&&
L_{i,j}(a)R_{i,j}(\varphi)=R_{i,j}(\varphi_{(1)})
\bra\varphi_{(2)}\,,\,a_{(1)}\ket L_{i,j}(a_{(2)})
\eea
}

\bsn{\it Proof:}
 We first use the left action (2.4) of $\A_{j+1}$ on $\A_{i,j}$
and the right action (2.5) of $\A_{i-1}$ on $\A_{i,j}$ to point out that the
assertions (4.13) and (4.14) are equivalent, respectively, to
\bealph
A_{i,j} \triangleleft A_{i-1} (a) &=&
L_{i,j} (S(a_{(1)}))A_{i,j} L_{i,j} (a_{(2)})\\
A_{j+1}(\varphi) \triangleright A_{i,j} &=&
R_{i,j} (\varphi_{(1)}) A_{i,j} R_{i,j}(S(\varphi_{(2)}))
\eealph
for all $A_{i-1}(a)\in\A_{i-1},\
A_{j+1}(\varphi)\in\A_{j+1}$ and $A_{i,j}\in\A_{i,j}$.
Note that equs. (4.16) say that these actions are inner in $\A_{i,j}$, as they
must be since $\A_{i,j}$ is simple for $j-i=2n+1$.

Given that $L_{i,j}$ commutes with $\A_{i+1,j}$ and $R_{i,j}$ commutes with
$\A_{i,j-1}$ eqns. (4.16) may also be rewritten as
\bealph
A_i(\psi)L_{i,j} (a) &=& L_{i,j} (a_{(1)})A_i(\psi\leftarrow a_{(2)})\\
R_{i,j} (\varphi) A_j(b) &=& A_j (\varphi_{(1)} \to b)R_{i,j} (\varphi_{(2)})
\eealph
To construct the maps $L_{i,j}$ and $R_{i,j}$
we now use the *-algebra isomorphism (2.12)
$$
\T_{i,j} :\A_{i,j} \to \A_{i,j-2} \otimes \End \H
$$
(assume without loss $\A_i\cong \hat H)$ and
proceed by induction over $n\in\NN_0$.
For $n=0$ we have $\T_{i,i+1}(\A_{i,i+1})=\End \H$, since
\bealph
\T_{i,i+1} (A_i(\psi)) &=& Q^+(\psi)\\
\T_{i,i+1} (A_{i+1}(b)) &=& P^+ (b)
\eealph
and we put
\bealph
L_{i,i+1}(a) &:=& T_{i,i+1}^{-1} \left(P^- (S^{-1}(a))\right)\\
R_{i,i+1} (\varphi) &:=& T_{i,i+1}^{-1}
\left(Q^-(S^{-1}(\varphi))\right)
\eealph
Then $L_{i,i+1}$ and $R_{i,i+1}$ define *-algebra
inclusions and (4.15) follows
straightforwardly from the definitions (2.7). Moreover,
$L_{i,i+1}(a)$ commutes
with $\A_{i+1}=\T_{i,i+1}^{-1}(P^+(H))$
and $R_{i,i+1}(\varphi)$ commutes with
$\A_i=\T_{i,i+1}^{-1} (Q^+(\hat H))$. Finally, using (4.18/19)
and (2.7) we get for $j=i+1$
$$
\ba{rcccl}
L_{i,i+1}(S(a_{(1)}))A_i(\psi)L_{i,i+1}(a_{(2)}) &=&
A_i(\psi\leftarrow a) &=& A_i(\psi)\triangleleft A_{i-1}(a)\\
R_{i,i+1}(\varphi_{(1)}) A_{i+1} (b) R_{i,i+1} (S(\varphi_{(2)})) &=&
A_{i+1} (\varphi\to b) &=& A_{i+2}(\varphi)\triangleright A_{i+1} (b)
\ea
$$
where the second equalities follow from (2.2), see also (4.2). This proves
(4.16) and therefore Proposition 4.5i)-iii) for $n=0$.

Assume now the claim holds for
$j=i+2n+1$ and put
\bealph
L_{i,j+2}(a) &:=& \T_{i,j+2}^{-1} \left(L_{i,j}(a) \otimes
\1\right)\\
R_{i,j+2}(\varphi) &:=& \T_{i,j+2}^{-1}
\left(R_{i,j}(\varphi_{(2)})\otimes Q^-(S^{-1}
(\varphi_{(1)}))\right)
\eealph
Then $L_{i,j+2}$ and $R_{i,j+2}$ again define *-algebra inclusions and (4.15)
immediately follows from the induction hypothesis. Also, since
$\T_{i,j+2}(\A_{j+1,j+2})=\1_\A \otimes \End \H$ we have
$$
L_{i,j+2}(a) \in\A_{i,j+2}\cap \A_{j+1,j+2}'
$$
Moreover, $\T_{i,j+2}(\A_{i+1,j})\subset \A_{i+1,j}\otimes P^-(H)$ commutes
with $L_{i,j}(a)\otimes\1$ by the induction hypothesis, and therefore
$L_{i,j+2}(a)\in\A_{i+1,j}'$ implies
\beq
L_{i,j+2}(a)\in\A_{i,j+2} \cap \A_{i+1,j+2}'.
\eeq
Next, to show that $R_{i,j}(\varphi)$ commutes with $\A_{i,j+1}$
we first note that $\T_{i,j+2}(\A_{i,j-1})=\A_{i,j-1}\o\1$ and
$\T_{i,j+2}(\A_{j+1 })=\1_\A\o Q^+(\hat H)$ and therefore
$$
R_{i,j+2}(\varphi)\in\A_{i,j+2} \cap \A_{i,j-1}' \cap \A_{j+1}'
$$
by (4.20b) and the induction hypothesis. To show that $R_{i,j+2}(\varphi)$
also commutes with $\A_j$ we compute
\beanon
\T_{i,j+2}(R_{i,j+2}(\varphi)A_j(b))
&=& R_{i,j}(\varphi_{(2)})A_j(b_{(1)})\otimes
Q^- (S^{-1} (\varphi_{(1)})) P^- (S(b_{(2)}))\\
&=& A_j (b_{(1)}) R_{i,j}(\varphi_{(3)})\otimes \bra \varphi_{(2)},b_{(2)}\ket
Q^-(S^{-1} (\varphi_{(1)}))P^-(S(b_{(3)}))\\
&=& A_j (b_{(1)})R_{i,j} (\varphi_{(2)})\otimes P^-(S(b_{(2)}))Q^- (S^{-1}
(\varphi_{(1)}))\\
&=& \T_{i,j+2}(A_j(b) R_{i,j+2} (\varphi))
\eeanon
where in the second line we have used the induction hypothesis in the form
(4.17b) and in the third line the Weyl algebra identity $P^-(b)Q^-(\varphi)=
Q^-(\varphi_{(2)})P^-(b_{(1)}) \bra \varphi_{(1)},b_{(2)} \ket$. Hence
$R_{i,j+2}(\varphi)$ also commutes with $\A_j$ and therefore
\beq
R_{i,j+2}(\varphi) \in\A_{i,j+2} \cap \A_{i,j+1}'
\eeq
To prove (4.13) for $L_{i,j+2}$ we note that
$\T_{i,j+2}=\T_{i-1,j+2}|\A_{i,j+2}$ and
$\T_{i-1,j+2}(A_{i-1}(a))=A_{i-1}(a)\otimes\1$, and therefore
\beanon
\T_{i-1,j+2}\left(A_{i-1}(a_{(1)})L_{i,j+2}(S(a_{(2)}))\right) &=&
A_{i-1}(a_{(1)})L_{i,j}(S(a_{(2)}))\o\1\\
&\in & (\A_{i-1,j}\cap\A'_{i,j})\o\1
\equiv\T_{i-1,j+2}(\A_{i-1,j+2}\cap\A'_{i,j+2})
\eeanon
by the induction hypothesis.
To prove (4.14) for $R_{i,j+2}$
we equivalently prove (4.17b) for $R_{i,j+2}$ by computing
\beanon
\T_{i,j+2}\left(R_{i,j+2}(\varphi)A_{j+2}(b)\right)
 &=& R_{i,j}(\varphi_{(2)}) \otimes
Q^-(S^{-1}(\varphi_{(1)}))P^+(b)\\
&=& R_{i,j}(\varphi_{(3)})\otimes P^+ (\varphi_{(1)} \to b)
Q^-(S^{-1}(\varphi_{(2)}))\\
&=& \T_{i,j+2}\left(A_{j+2}(\varphi_{(1)}\to b)
R_{i,j+2}(\varphi_{(2)})\right)
\eeanon
where the Weyl algebra identity used in the second line
follows again straightforwardly
from (2.7). This concludes the proof of Proposition 4.5.
\qed

\bsn
As a particular consequence of Proposition 4.5 we also need

\bsn
{\bf Corollary 4.6:}
{\sl For all $A_j(a)\in \A_j$ and $A_{j+1}(\varphi) \in
\A_{j+1}$ we have}
\bea
i)~~~ &A_{j+1}(S(\varphi_{(1)})) R_{i,j} (\varphi_{(2)})
 = R_{i,j}(\varphi_{(2)})
A_{j+1}(S(\varphi_{(1)}))\in \A_{i,j+1} \cap \A_{i,j}'\\
ii)~~~& R_{i,j}(\varphi)A_j(a)
= A_j(a_{(1)}) R_{i,j}(\varphi\leftarrow a_{(2)})
\eea

\bsn
{\it Proof:}
\beanon
i) ~~~A_{j+1} (S(\varphi_{(1)}))R_{i,j}(\varphi_{(2)})
&=& R_{i,j}\left(S(S(\varphi_{(2)})\varphi_{(3)})\right)
A_{j+1} (S(\varphi_{(1)}))R_{i,j}(\varphi_{(4)})\\
&=&R_{i,j}(S^2(\varphi_{(2)})) A_{j+1}(S(\varphi_{(1)}))R_{i,j} (S(\varphi_{(3)})
\varphi_{(4)})\\
&=& R_{i,j}(\varphi_{(2)})A_{j+1} (S(\varphi_{(1)}))\in\A_{i,j+1} \cap \A_{i,j}'
\eeanon
where in the second line we have used (4.14) and in last line
$S^2=\id$.
\beanon
ii)~~~ R_{i,j}(\varphi)A_j(a) &=& A_{j+1}(\varphi_{(1)} S(\varphi_{(2)}))
R_{i,j} (\varphi_{(3)})A_j(a)\\
&=& A_{j+1}(\varphi_{(1)})A_j(a)A_{j+1}
(S(\varphi_{(2)}))R_{i,j}(\varphi_{(3)})\\
&=& A_j (a_{(1)} ) A_{j+1}(\varphi_{(1)} \leftarrow a_{(2)}) A_{j+1}(S(\varphi_{(2)}))
R_{i,j}(\varphi_{(3)})\\
&=& A_j(a_{(1)})R_{i,j}(\varphi\leftarrow a_{(2)})
\eeanon
where in the second line we have used (4.23) and the the third line (2.2b).
\qed

\bigskip
Using Proposition 4.5 and Corollary 4.6 we are now in the
position to prove Theorem 3.12 as a particular consequence of
the following

\bsn{\bf Theorem 4.7:}
{\sl Let $j=i+2n+1,\ n\in\NN_0,\ i\in\ZZ$, and let
$I=[i-\half,j+\half]\in\I$. Define $\r_{i-1,j+1}:\A(\partial
I)\to\A_{i-1,j+1}\o\D(H)$ by
\bealph
\r_{i-1,j+1}(A_{j+1}(\varphi)) &:=& R_{i,j}(\varphi_{(1)}S(\varphi_{(3)}))
A_{j+1}(\varphi_{(4)})\o\D(\varphi_{(2)})\\
\r_{i-1,j+1}(A_{i-1}(a)) &:=&
A_{i-1}(a_{(1)})L_{i,j}(S(a_{(2)})a_{(4)})\o\D(a_{(3)})
\eealph
Then
\bit
\item[i)] $\r_{i-1,j+1}$ extends to a coaction
  $\hat\r_{i-1,j+1}\in\Amp(\A,\partial I)$, which is strictly equivalent
  to $\r_{i-1,i}$.
\item[ii)] The coaction $\hat\r_{i-1,j+1}$ is universal in
$\Amp(\A,\partial I)$.
\eit
}

\bsn {\it Proof:}
 Assume without loss $\A_i\simeq \hat H$ and define
\bealph
T_{i,j} :=\sum_k L_{i,j} (b_k) \otimes \D(\xi^k) \in\A_{i,j}\otimes\D(H)
\eealph
where $b_k\in H$ is a basis with dual basis $\xi^k\in\hat H$. Then
$T_{i,j}$ is unitary,
$$
T_{i,j}^* =T_{i,j}^{-1} =\sum_k L_{i,j} (b_k) \otimes \D(S(\xi^k))
\eqno (4.26b)
$$
and we put
\beq
\hat\rho_{i-1,j+1} := \Ad T_{i,j} \circ \rho_{i-1,i}
\eeq
To prove i) we first show
\beq
\hat\rho_{i-1,j+1} \in \Amp(\A,\partial I)
\eeq
and
\beq
\hat\rho_{i-1,j+1}| \A(\partial I)=\rho_{i-1,j+1} .
\eeq
To this end we use that $L_{i,j}(a) \in\A_{i,j}\cap \A_{i+1,j}'$ to conclude
$$
T_{i,j} \in(\A_{-\infty,i-2}' \cap \A_{i+1,j}' \cap \A_{j+2,\infty}')
\otimes \D(H)
$$
Now $\A((\partial I)^c)=\A_{-\infty,i-2} \lor \A_{i,j} \lor \A_{j+2,\infty}$
and since $\rho_{i-1,i}$ is localized on $\A_{i-1,i}$ the claim (4.28) follows
provided
\beq
(A_i(\varphi)\otimes\1)\,T_{i,j} = T_{i,j}\, \rho_{i-1,i} (A_i(\varphi)),
\quad \forall \varphi \in\hat H.
\eeq
To check (4.30) we compute
\beanon
(A_i(\varphi)\otimes \1)T_{i,j} &=& \sum_k A_i(\varphi)L_{i,j}(b_k)\otimes
\D(\xi^k)\\
&=& \sum_{k_1,k_2}L_{i,j}(b_{k_1})A_i(\varphi\leftarrow b_{k_2})\otimes
\D(\xi^{k_1}\xi^{k_2})\\
&=& \sum_k L_{i,j}(b_k)A_i(\varphi_{(2)}) \otimes \D(\xi^k\varphi_{(1)})\\
&=& T_{i,j} \rho_{i-1,i} (A_i(\varphi))
\eeanon
where in the second line we have used (4.17a).
Thus we have proven (4.28). To prove
(4.29) we compute
\bealph
&&\rho_{i-1,j+1} (A_{j+1}(\varphi)) T_{i,j}=\nonumber\\
&&\qquad= \sum_k R_{i,j} (\varphi_{(1)}
S(\varphi_{(3)}))A_{j+1} (\varphi_{(4)})L_{i,j} (b_k)\o \D(\varphi_{(2)}
\xi^k)\\
&&\qquad= \sum_k R_{i,j}(\varphi_{(1)})L_{i,j}(b_k)R_{i,j} (S(\varphi_{(3)}))
A_{j+1}(\varphi_{(4)} ) \otimes \D(\varphi_{(2)}\xi^k)\nonumber\\
&&\qquad=\sum_{k_1,k_2} L_{i,j}(b_{k_2})
R_{i,j}(S^{-1} (b_{k_1})\to \varphi_{(1)})
R_{i,j}(S(\varphi_{(3)}))A_{j+1}(\varphi_{(4)})\otimes \D
(\varphi_{(2)} \xi^{k_1} \xi^{k_2})\nonumber\\
&&\qquad= \sum_k L_{i,j} (b_k)R_{i,j}
(\varphi_{(1)}S(\varphi_{(4)}))A_{j+1}(\varphi_{(5)})
\otimes \D(\varphi_{(3)} S^{-1} (\varphi_{(2)})\xi^k)\nonumber\\
&&\qquad= \sum_k L_{i,j} (b_k) A_{j+1} (\varphi)\otimes \D(\xi^k)\nonumber\\
&&\qquad= T_{i,j} \,(A_{j+1} (\varphi)\otimes \1)\\
&&\qquad= T_{i,j}\, \rho_{i-1,i} (A_{j+1}(\varphi))
\eealph
where in the second equation we have used (4.14) and in the
third equation the inverse of (4.15).
Next we compute
\beanon
T_{i,j}\,\rho_{i-1,i} (A_{i-1}(a)) &=& T_{i,j}\,[A_{i-1}(a_{(1)})\otimes \D
(a_{(2)})]\\
&=& T_{i,j}\,
[A_{i-1}(a_{(1)})L_{i,j} (S(a_{(2)}) a_{(3)}) \otimes
\D(a_{(4)})]\\
&=& [A_{i-1} (a_{(1)})L_{i,j}(S(a_{(2)})) \otimes \1]\,
 T_{i,j}\, [L_{i,j}(a_{(3)})
\otimes \D(a_{(4)})]\\
&=&[A_{i-1}(a_{(1)} )L_{i,j}(S(a_{(2)})a_{(4)})
 \otimes \D(a_{(3)})] \,T_{i,j}\\
&=& \rho_{i-1,j+1} (A_{i-1} (a))\, T_{i,j}
\eeanon
where in
the third line we have used (4.13) and in the fourth line the identity
\beq
T_{i,j}\,[L_{i,j}(a_{(1)})\otimes \D(a_{(2)}] =[L_{i,j}(a_{(2)})
 \otimes \D(a_{(1)})]\, T_{i,j}
\eeq
which follows straightforwardly from equ. (B.2) in Appendix B.
Thus we have proven (4.29).
To complete the proof of part i)
we are left to show that $\r_{i-1,j+1}$ provides a coaction
which is strictly equivalent to $\r_{i-1,i}$. This follows
provided
\beq
T_{i,j}\x_{\rho_{i-1,i}} T_{i,j} =(id\otimes\Delta_\D^{(op)})(T_{i,j})
\eeq
To prove  (4.33) we use that $L_{i,j}(b_k)$ lies in $\A_{i,j}$ and therefore
$(\hat\r_{i-1,j+1}\o\id)(T_{i,j})=T_{i,j}^{02}$ implying
\beanon
T_{i,j}\x_{\rho_{i-1,i}} T_{i,j} &=&(\hat\r_{i-1,j+1}\o\id)(T_{i,j})
(T_{i,j}\o\1)\\
&=& T_{i,j}^{02}T_{i,j}^{01}\\
&=&(\id\o\Delta_\D^{(op)})(T_{i,j})
\eeanon
Thus we have proven part i) of Theorem 4.7.

To prove part ii) first recall
that $\rho_{i-1,i}$ is effective and therefore
$\hat\rho_{i-1,j+1}=\Ad T_{i,j} \circ \rho_{i-1,i}$ is effective. Let now
$\mu\in\Amp(\A,\partial I)$
and define $\hat\mu :\A_{j+1} \to \A\otimes \End V_\mu$ by
\bealph
\hat \mu(A_{j+1}(\varphi))
:=\mu(A_{j+1}(\varphi_{(2)}))[A_{j+1}(S(\varphi_{(3)}))
R_{i,j}(\varphi_{(4)}S^{-1}(\varphi_{(1)}))\otimes \1]
\eealph
Then $\mu$ may be expressed in terms of $\hat\mu$
$$
\ba{rcl}
 \mu(A_{j+1}(\varphi)) &=&
\mu (A_{j+1}(\varphi_{(3)}))\, [R_{i,j}(S^{-1}(\varphi_{(2)})) A_{j+1}
(S(\varphi_{(4)}))R_{i,j}(\varphi_{(5)})\otimes\1]\\
&&\x[R_{i,j}(S(\varphi_{(6)}))A_{j+1}
(\varphi_{(7)})R_{i,j}(\varphi_{(1)})\otimes\1] \\
&=&\hat\mu (A_{j+1}(\varphi_{(2)}))\,[R_{i,j} (\varphi_{(1)}
S(\varphi_{(3)}))A_{j+1} (\varphi_{(4)}) \otimes \1]
\ea
\eqno (4.34b)
$$
where in the second equation we have used (4.14).
In Lemma 4.8 below we show
that there exists a *-representation $\beta_\mu:\hat H\to\End V_\mu$ such that
\beq
\hat\mu(A_{j+1}(\varphi))=\1_\A \otimes\beta_\mu(\varphi)
\eeq
Then (4.34b) implies
\beq
\mu(A_{j+1}(\varphi))=R_{i,j}(\varphi_{(1)}
S(\varphi_{(3)}))A_{j+1} (\varphi_{(4)})
\otimes \beta_\mu(\varphi_{(2)})\quad.
\eeq
Putting
\beq
V_{i,j}=\sum_kL_{i,j}(b_k)\otimes \beta_\mu(\xi^k)
\eeq
and repreating the calculation from (4.31a) to (4.31b)
with $\rho_{i-1,j+1}$
replaced by $\mu, \
T_{i,j}$ replaced by $V_{i,j}$ and $\D(\varphi)$ replaced
by $\beta_\mu(\varphi)$ we get
\beq
\mu(A_{j+1}(\varphi))V_{i,j} = V_{i,j}
(A_{j+1}(\varphi)\otimes\1).
\eeq
Moreover, similarly as for $T_{i,j}$ we have
\beq
V_{i,j} \in (\A'_{-\infty,i-2}
\cap \A'_{i+1,j}\cap \A'_{j+2,\infty}) \otimes\End V_\mu\quad.
\eeq
By (4.38) and (4.39)
$\Ad V^*_{i,j} \circ \mu$ is localized on $\A_{i-1,i}$.
In particular
\beq
V_{i,j}^* \,\mu(A_i(\varphi))V_{i,j}
\equiv V_{i,j}^* (A_i(\varphi)\otimes \1)
V_{i,j} =A(\varphi_{(2)})\otimes \beta_\mu(\varphi_{(1)})
\eeq
which one proves in the same way as (4.30). Hence, by Theorem 4.1ii)
$\beta_\mu$ extends to a represenation $\hat\beta_\mu:\D(H)\to\End V_\mu$
such that
$$
\Ad V_{i,j}^*\circ \mu=(id\otimes \hat\beta_\mu)\circ \rho_{i-1,i}
$$
and therefore
\beq
\mu=(id\otimes\hat\beta_\mu)\circ \rho_{i-1,j+1}.
\eeq
This proves that $\rho_{i-1,j+1}$ is universal in
$\Amp (\A,\partial I)$ and therefore part ii) of Theorem 4.7.
\qed

\bigskip
Since by Proposition 4.2 the coactions $\r_{i-1,i},\ i\in\ZZ$,
are all (cocycle) equivalent and since by Corollary 3.9 any
amplimorphism $\mu\in\Amp\A$ is compressible into $\partial I$
for some interval $I\in\I$ of even length, Theorem 4.7 implies
that $\Amp\A$ is compressible into {\em any} interval of length
two. In particular, $\Amp\A$ is completely compressible.
This concludes the proof of Theorem 3.12.

\smallskip
We are left to prove the claim (4.35).

\bsn
{\bf Lemma 4.8:}
{\sl Under the conditions of Theorem 4.7 let $\mu\in\Amp (\A,
\partial I)$ and let $\hat \mu:\A_{j+1}\to \A_{i,j+1}\otimes
\End V_\mu$ be given by (4.34a).
Then there exists a *-representation $\beta_\mu:
\A_{j+1}\to \End V_\mu$ such that $\hat\mu=\1_\A\otimes\beta_\mu$.
}

\bsn
{\it Proof:}
Since $\partial I\subset \overline I$ we have by Lemma 3.8
$$
\mu(\A(\partial I))\subset \A_{i-1,j+1}\otimes \End V_\mu
$$
Using $\A_{j+1}\subset \A(\partial I)\cap \A'_{i-2} \cap \A'_{i,j-1}$
we conclude
\beanon
\mu(\A_j) &\subset &
(\A_{i-1,j+1} \otimes \End V_\mu)\cap\mu(\A_{i-2})' \cap
\mu (\A_{i,j-1})' \\
&=& (\A_{i-1,j+1} \cap\A_{i-2}' \cap \A'_{i,j-1} )\otimes \End V_\mu\\
 &=& (\A_{i,j+1} \cap \A'_{i,j-1})\otimes \End V_\mu
\eeanon
Let now
\beq
\lambda(\varphi) := \mu(A_{j+1} (\varphi_{(1)})) [A_{j+1} (S(\varphi_{(2)}))
\otimes\1]
\eeq
Using that $\mu|\A_{j+2}=\id\otimes \1$ we conclude
\beanon
[A_{j+2}(a)\otimes\1] \lambda (\varphi)
&=& \mu(A_{j+1}(a_{(1)}\to \varphi_{(1)}))
[A_{j+1}(a_{(2)}\to S(\varphi_{(2)})) A_{j+2}(a_{(3)}) \otimes \1]\\
&=& \mu(A_{j+1}(\varphi_{(1)})) [A_{j+1}(S(\varphi_{(4)})) A_{j+2} (a_{(2)})
\bra a_{(1)},\varphi_{(2)} S(\varphi_{(3)})\ket \otimes\1]\\
&=& \lambda(\varphi) [A_{j+2}(a)\otimes\1]
\eeanon
and therefore
\beanon
\lambda(\varphi)\in(\A_{i,j+1} \cap \A'_{j+2} \cap \A'_{i,j-1}) \otimes \End V_\mu\\
=(\A_{i,j} \cap\A'_{i,j-1})\otimes \End V_\mu
\eeanon
Thus we get
\bea
\hat\mu(\varphi) &\equiv&
 \lambda(\varphi_{(2)})[R_{i,j}(\varphi_{(3)} S^{-1}
(\varphi_{(1)}))\otimes \1]\nonumber\\
&\in& (\A_{i,j} \cap \A'_{i,j-1})\otimes \End V_\mu
\eea
We claim  that $\hat\mu(\varphi)$
commutes with $\A_j\otimes\1$ and therefore
\bea
\hat\mu(\varphi) &\in&
(\A_{i,j} \cap \A'_{i,j}) \otimes \End V_\mu\nonumber\\
&=&\1_\A \otimes \End V_\mu
\eea
by the simplicity of $\A_{i,j}$.
To this end we use (4.23) and (4.24) and $\mu
(A_j(a))=A_j(a)\otimes\1$ to compute
\bea
&&\hat\mu (\varphi)\,[A_j(a)\otimes\1]=\nonumber\\
&&=\mu(A_{j+1}(\varphi_{(2)}))\,
[R_{i,j}(S^{-1}(\varphi_{(1)})) A_j(a) A_{j+1}(S(\varphi_{(3)}))
R_{i,j} (\varphi_{(4)})\otimes\1]\nonumber\\
&&=[A_j(a_{(1)})\otimes\1]\,
 \mu(A_{j+1}(\varphi_{(2)} \leftarrow a_{(2)}))\,
[R_{i,j} (S^{-1}(\varphi_{(1)})\leftarrow a_{(3)})A_{j+1} (S(\varphi_{(3)}))
R_{i,j} (\varphi_{(4)})\otimes \1]\nonumber\\
&&= [A_j(a_{(1)})\,\bra a_{(2)},\varphi_{(3)} S^{-1}(\varphi_{(2)})\ket
\otimes \1] \,\mu(A_{j+1} (\varphi_{(4)}))\,
[A_{j+1} (S(\varphi_{(5)}))R_{i,j} (\varphi_{(6)} S^{-1}(\varphi_{(1)}))
\otimes\1]\nonumber\\
&&= [A_j (a)\otimes \1]\, \hat\mu (\varphi).
\eea
From (4.43) and (4.45) we get (4.44) and therefore
$$
\hat\mu(\varphi) =\1_\A\otimes \beta_\mu(\varphi)
$$
for some linear map $\beta_\mu:\A_{j+1}\to \End V_\mu$. We are left to check
that $\beta_\mu$ provides a *-representation:
\beanon
\hat\mu (\varphi)\hat\mu(\psi) &=&
(A_{j+1}(\varphi_{(2)}))\,\hat\mu(\psi)\,
[A_{j+1}(S(\varphi_{(3)}))
R_{i,j} (\varphi_{(4)}S^{-1}(\varphi_{(1)}))\otimes\1]
\\
&=& \mu(A_{j+1}(\varphi_{(2)}\psi_{(2)})\,
[A_{j+1}(S(\varphi_{(3)}\psi_{(3)}))R_{i,j}(\varphi_{(4)}
\psi_{(4)} S^{-1} (\psi_{(1)})S^{-1}(\varphi_{(1)})\otimes\1]
\\
&=&\hat\mu(\varphi\psi)
\eeanon
where in the second line we have used (4.23).
\beanon
\hat\mu(\psi^*)^*
&=& [R_{i,j} (S(\psi_{(1)})\psi_{(4)})A_{j+1}(S^{-1}(\psi_{(3)}))
\otimes \1]\, \mu(A_{j+1}(\psi_{(2)}))
\\
&=& R_{i,j} (S(\psi_{(1)})\psi_{(7)})A_{j+1} (S^{-1}(\psi_{(6)})) R_{i,j}
(\psi_{(2)} S(\psi_{(4)}))A_{j+1}(\psi_{(5)})\otimes \beta_\mu
(\psi_{(3)})
\\
&=& R_{i,j} (S(\psi_{(1)})\psi_{(2)}S(\psi_{(4)}) \psi_{(7)}) A_{j+1}
(S^{-1} (\psi_{(6)})\psi_{(5)}) \otimes \beta_\mu(\psi_{(3)})
\\
&=& \1 \otimes \beta_\mu(\psi)
\eeanon
where in the second line we have used (4.36) and in the third line (4.14).
\qed

\newpage
\begin{appendix}

\section{Finite dimensional C$^*$-Hopf algebras}

There is an extended literature on Hopf algebra theory the
nomenclature of which, however, is by far not unanimous
[BaSk,Dr,E,ES,Sw,W]. Therefore we
summarize in this appendix some standard notions in order to fix
our conventions and notations.

A linear space $B$ over $\CC$ together with linear maps
$$
\ba{rcl}
m\colon B\o B&\to& B\quad \hbox{(multiplication)},\\
\iota\colon\CC&\to& B\quad\hbox{(unit)},
\ea
\ba{rcl}
\Delta\colon B&\to& B\o
B\quad\hbox{(comultiplication)},\\
\varepsilon\colon B&\to&\CC\quad\hbox{(counit)}
\ea
$$
is called a {\em bialgebra } and denoted by $B(m,\iota,\Delta,\varepsilon)$
if the following axioms hold:
\beanon
&
\ba{rcl}
  m\circ(m\o\id)&=&m\circ(\id\o m)\,,\\
  m\circ(\iota\o\id)&=&m\circ(\id\o\iota)=\id\,,\\
  \varepsilon\circ m &=& \varepsilon\o\varepsilon\,,
\ea
\ba{rcl}
   (\Delta\o\id)\circ\Delta&=&(\id\o\Delta)\circ\Delta\\
   (\varepsilon\o\id)\circ\Delta&=&(\id\o\varepsilon)\circ\Delta=\id\\
   \Delta\circ\iota=\iota\o\iota
\ea
&\\
&\Delta\circ m=(m\o m)\circ\tau_{23}\circ(\Delta\o\Delta)&
\eeanon
where $\tau_{23}$ denotes the permutation of the tensor factors
2 and 3.
We use Sweedler's notation $\Delta(x)=x_{(1)}\o x_{(2)}$, where
the right hand side is understood as a sum $\sum_i
x_{(1)}^i\o x_{(2)}^i\in B\o B$. For iterated coproducts we
write $x_{(1)}\o x_{(2)}\o x_{(3)}:=\Delta(x_{(1)})\o
x_{(2)}\equiv x_{(1)}\o\Delta(x_{(2)})$, etc.
The image under $\iota$ of the number $1\in\CC$ is the unit element of
$B$ denoted by $\one$. The linear dual $\hat B$ becomes also a
bialgebra by transposing the structural maps
$m,\iota,\Delta,\varepsilon$ by means of the canonical pairing
$\langle\ \ ,\ \ \rangle\colon \hat B\times B\to\CC$.

A bialgebra $H(m,\iota,\Delta,\varepsilon)$ is called a {\em Hopf
algebra } $H(m,\iota,S,\Delta,\varepsilon)$ if there exists an
antipode $S\colon H\to H$, i.e. a linear map satisfying
$$
m\circ(S\o\id)\circ\Delta=m\circ(\id\o S)\circ\Delta=
\iota\circ\varepsilon
\eqno(A.1)$$
Using the above notation equ. (A1) takes the form
$S(x_{(1)})x_{(2)}=x_{(1)}S(x_{(2)})=\varepsilon(x)\one$,
which in connection with the coassociativity of $\Delta$ is
often applied in formulas involving iterated coproducts like,
e.g., $x_{(1)}\o x_{(4)}S(x_{(2)})x_{(3)}=x_{(1)}\o x_{(2)}$.
All other properties of the antipode, i.e. $S(xy)=S(y)S(x),\
\Delta\circ S=(S\o S)\circ\Delta_{op}$ and
$\varepsilon\circ S=\varepsilon$, as well as the uniqueness of
$S$ are all consequences of the
axiom (A.1) [Sw].
The dual bialgebra $\hat H$ of $H$ is also a Hopf algebra with the
antipode defined by
$$
\langle S(\varphi),x\rangle:=
\langle\varphi,S(x)\rangle\quad\varphi\in\hat H,\ x\in H\,.
\eqno(A.2)
$$
A $*$-Hopf algebra $H(m,\iota,S,\Delta,\varepsilon,*)$ is a Hopf
algebra $H(m,\iota,S,\Delta,\varepsilon)$ together with an antilinear
involution $^*\colon H\to H$ such that $H(m,\iota,*)$ is a
$*$-algebra and $\Delta$ and $\varepsilon$ are $^*$-algebra
maps.
It follows that $\overline S:=*\circ S\circ *$ is the antipode
in the Hopf algebra
$H_{op}$ (i.e. with opposite muliplication) and therefore
$\overline S=S^{-1}$ [Sw].
The dual of a $*$-Hopf algebra is also a $*$-Hopf algebra with
$^*$-operation defined by $\varphi^*:=S(\varphi_*)$, where
$\varphi\mapsto\varphi_*$ is the antilinear involutive algebra
automorphism given by
$$
\langle
\varphi_*,x\rangle:=\overline{\langle\varphi,x^*\rangle}\,.
\eqno(A.3)
$$

Let $\A$ be a $*$-algebra and let $H$ be a $*$-Hopf algebra. A (Hopf
module) left
action of $H$ on $\A$ is a linear map $\gamma\colon H\o \A\to \A$
satisfying the following axioms: For $A,B\in \A$, $x,y\in H$
$$
\ba{rcl}
\gamma_x\circ\gamma_y(A)&=&\gamma_{xy}(A)\\
\gamma_x(AB)&=&\gamma_{x_{(1)}}(A)\gamma_{x_{(2)}}(B)\\
\gamma_x(A)^*&=&\gamma_{x_*}(A^*)
\ea
\eqno(A.4)
$$
where as above $x_*=S^{-1}(x^*)$.
A right action of $H$ is a left action of $H_{op}$.
Important examples are the action of $H$ on $\hat H$ and that of
$\hat H$ on $H$ given by the Sweedler's arrows:
\setc{4}
\bealph
\gamma_x(\varphi)=x\to\varphi&:=&\varphi_{(1)}\langle
x,\varphi_{(2)}\rangle\\
\gamma_{\varphi}(x)=\varphi\to x&:=&
x_{(1)}\langle\varphi,x_{(2)}\rangle
\eealph
A left action is called inner if there exists a *-algebra map
$i:H\to\A$ such that
$\gamma_x(A)=i(x_{(1)})\,A\,i(S(x_{(2)}))$. Left $H$-actions
$\gamma$ are in one-to-one corespondence with right {\em $\hat
H$-coactions} (often denoted by the same symbol)
$\gamma:\A\to\A\o \hat H$ defined by
$$
\gamma(A):=\gamma_{b_i}(A)\o\xi^i,\quad A\in\A
$$
where $\{b_i\} $ is a basis in $H$ and $\{\xi^i\}$ is the dual
basis in $\hat H$ and where for simplicity we assume from now
on $H$ to be finite dimensional. Conversely, we have
$\gamma_x=(\idA\o x)\circ\gamma$. The defining properties of a
coaction are given in equs. (3.11a-e).

Given a left $H$-action (right $\hat H$-coaction) $\gamma$ one
defines the {\em crossed product}
$\A\cros_\gamma H$ as the $\CC$-vector space $\A\o H$ with
$*$-algebra structure
\bealph
(A\o x)(B\o y) &:=& A\gamma_{x_{(1)}}(B)\o x_{(2)}y \\
(A\o x )^* &:=& (\1_\A\o x^*)(A^*\o\1_H)
\eealph
An important example is the "Weyl algebra" $\W(\hat H):=\hat
H\cros H$, where the crossed product is taken with
respect to the natural left action (A.5a). We have $\W(\hat
H)\cong\End\hat H$ where the isomorphism is given by (see
[N] for a review)
\beq
w:\ \psi\o x\mapsto Q^+(\psi)P^+(x)\ .
\eeq
Here we have introduced $Q^+(\psi),\ \psi\in\hat H$ and
$P^+(x),\ x\in H$ as operators in $\End\hat H$ defined on
$\xi\in\hat H$ by
\beanon
Q^+(\psi)\xi &:=& \psi\xi\\
P^+(x)\xi &:=& x\to\xi
\eeanon

Any right $H$-coaction $\beta\,:\A\to\A\o H$ gives rise to a natural
left $H$-action $\gamma$ on $\A\cros_\beta\hat H$
\beq
\gamma_x(A\o\psi):=A\o(x\to\psi)
\eeq
The resulting iterated crossed product $(\A\cros_\beta\hat
H)\cros_\gamma H$ contains $\W(\hat H)\cong\End\hat H$ as the
subalgebra given by $\1_\A\o\psi\o x\cong Q^+(\psi)P^+(x),\
\psi\in\hat H,\ x\in H$. Moreover, by the Takesaki duality
theorem [Ta,NaTa] the iterated crossed product $(\A\cros_\beta\hat H)
\cros_\gamma H$ is canonically isomorphic to $\A\o\End \hat H$. In fact,
defining the representation $L:H\to\End\hat H$ by
\beq
L(x)\xi:=\xi\leftarrow
S^{-1}(x)\equiv\bra\xi_{(1)}\,,\,S^{-1}(x)\ket\xi_{(2)}
\eeq
one easily verifies that $\T:(\A\cros_\beta\hat H)\cros_\gamma H
\to\A\o\End\hat H$
\bealph
\T(A\o\1_{\hat H}\o\1_H)&:=&(\idA\o L)(\beta(A))\\
\T(\1_\A\o\psi\o x)&:=&\1_\A\o Q^+(\psi)P^+(x)
\eealph
defines a $*$-algebra map. $\T$ is surjective since $w$ is
surjective and therefore $\1_\A\o\End\hat H\subset\Im\T$ and
\beanon
A\o\1_{\End\hat H}&\equiv& A_{(0)}\o L(A_{(1)}S(A_{(2)}))\\
&=&\T(A_{(0)}\o\1_{\hat H }\o\1_H)(\1_\A\o L(S(A_{(1)})))\\
&\in&\Im\T
\eeanon
for all $A\in\A$.
Here we have used the notation
$A_{(0)}\o A_{(1)}=\beta(A)$,
$$
A_{(0)}\o A_{(1)}\o A_{(2)}=(\beta\o\id_H)(\beta(A))\equiv
(\idA\o\Delta)(\beta(A))
$$
(including a summation convention) and the identity
$(\idA\o\varepsilon)\circ\beta=\idA$,
see equs. (3.11d,e).
The inverse of $\T$ is given by
\bealph
\T^{-1}(\1_A\o W)&=&\1_\A\o w^{-1}(W)\\
\T^{-1}(A\o\1_{\End\hat H}) &=& A_{(0)}\o w^{-1}(L(S(A_{(1)})))
\eealph
for $W\in\End\hat H$ and $A\in\A$.

\bigskip
A {\em left(right) integral} in $\hat H$
is an element $\chi^L(\chi^R)\in\hat H$ satisfying
\bealph
\varphi\chi^L=\varepsilon(\varphi)\chi^L
\qquad\chi^R\varphi=\varepsilon(\varphi)\chi^R
\eealph
for all $\varphi\in\hat H$ or equivalently
$$
\chi^L\rightarrow x=\langle\chi^L,x\rangle\one\,,\qquad
  x\leftarrow\chi^R=\langle\chi^R,x\rangle\one
\eqno(A.12b)
$$
for all $x\in H$. Similarly one defines left(right) integrals
in $H$.

If $H$ is finite dimensional and semisimple then so is
$\hat H$ [LaRa] and in this case they are both {\em unimodular},
i.e. left and right integrals coincide and are all given as
scalar multiples of a unique one dimensional central projection
\beq
e_\varepsilon=e_\varepsilon^*=e_\varepsilon^2=S(e_\varepsilon)
\eeq
which is then called the {\em Haar integral}.

For $\varphi,\psi\in\hat H$ and $h\equiv e_\varepsilon\in H$
the Haar integral define the hermitian form
\beq
\bra\varphi|\psi\ket:=\bra\varphi^*\psi,h\ket
\eeq
Then $\bra\cdot|\cdot\ket$ is nondegenerate [LaSw] and it is
positve definite --- i.e. the Haar integral
$h$ provides a positive state ({\em the Haar "measure"}) on $\hat H$
--- if and only if $\hat H$ is a {\em $C^*$-Hopf
algebra}.
These are the "finite matrix pseudogroups" of [W].
They also satisfy $S^2=\id$ and $\Delta(h)=\Delta_{op}(h)$ [W].
If $\hat H$ is a finite dimensional $C^*$-Hopf algebra then so is
$H$, since $H\ni x\to P^+(x)\in\End\hat H$ defines a faithful
$*$-representation on the Hilbert space $\H\equiv L^2(\hat H,h)$.
Hence finite dimensional $C^*$-Hopf algebras always come in
dual pairs. Any such pair serves as a building block for our Hopf spin
model.

\sec{The Drinfeld Double}

Here we list the basic properties
of the Drinfeld double $\D(H)$ (also called quantum double)
of a finite dimensional $*$-Hopf
algebra $H$ [Dr,Maj1]. Although most of them are well known in the
literature, the presentation (B.1) by generators and relations
given below seems to be new.

As a $*$-algebra $\D(H)$ is generated by elements $\D(a),\ a\in
H$ and $\D(\varphi),\ \varphi\in\hat H$ subjected to the following
relations:
\bealph
\D(a)\D(b)&=&\D(ab)
\\
\D(\varphi)\D(\psi)&=&\D(\varphi \psi)
\\
\D(a_{(1)})\,\langle
a_{(2)},\varphi_{(1)}\rangle\,\D(\varphi_{(2)})&=&
\D(\varphi_{(1)})\,\langle
\varphi_{(2)},a_{(1)}\rangle\,\D(a_{(2)})
\\
\D(a)^*=\D(a^*) &,& \D(\varphi)^*=\D(\varphi^*)
\eealph
The relation (B.1c) is equivalent to any one of the following
two relations
\bealph
\D(a)\D(\varphi)&=&\D(\varphi_{(2)})\D(a_{(2)})\,
  \langle a_{(1)},\varphi_{(3)}\rangle
  \langle S^{-1}(a_{(3)}),\varphi_{(1)}\rangle
\\
\D(\varphi)\D(a)&=&\D(a_{(2)})\D(\varphi_{(2)})\,
  \langle \varphi_{(1)},a_{(3)}\rangle
  \langle S^{-1}(\varphi_{(3)}),a_{(1)}\rangle
\eealph
These imply that as a linear space $\D(H)\cong H\o\hat H$
and also that as a $*$-algebra $\D(H)$ and $\D(\hat H)$ are
isomorphic. This $*$-algebra will be denoted by $\G$.

The Hopf algebraic structure of $\D(H)$ is given by the following
coproduct, counit, and antipode:
\bealph
\Del(\D(a))=\D(a_{(1)})\o \D(a_{(2)}) &\quad&
\Del(\D(\varphi))=\D(\varphi_{(2)})\o \D(\varphi_{(1)})
\\
\varepsilon_{\D}(\D(a))=\varepsilon(a) &\quad&
\varepsilon_{\D}(\D(\varphi))=\varepsilon(\varphi)
\\
S_{\D}(\D(a))=\D(S(a)) &\quad&
S_{\D}(\D(\varphi))=\D(S^{-1}(\varphi))
\eealph
It is straightforward to check that equs. (B.3) provide a $*$-Hopf algebra
structure on  $\D(H)$. Moreover,  $\D(\hat H)= (\D(H))_{cop}$
(i.e. with opposite coproduct) by (B.3a).

If $H$ and $\hat H$ are $C^*$-Hopf algebras then so is $\D(H)$.
To see this one may
use the faithful $*$-representations of  $\D(H)$ on the Hilbert
spaces $\H_{n,m}$ in Lemma 2.2.
Alternatively, it is not difficult to see that
\beq
\D(h)\D(\chi)=\D(\chi)\D(h)=:h_\D
\eeq
provides the Haar integral in  $\D(H)$ and that the positivity
of the Haar states $h\in H$ and $\chi\in\hat H$ implies the positvity of the
state $h_\D$ on $\widehat{\D(H)}$ .

The dual $\widehat{\D(H)}$ of $\D(H)$ has been studied by [PoWo].
As a coalgebra it is $\hat \G$ and coincides
with the coalgebra $\widehat{\D(\hat H)}$. The latter one,
however, as an algebra differs
from $\widehat{\D(H)}$ in that the multiplication is replaced
by the opposite multiplication.

\medskip
The remarkable property of the double construction is that it
always yields a {\em quasitriangular} Hopf algebra [Dr].
By definition this means that there exists a unitary $R\in \D(H)\o
\D(H)$ satisfying the hexagonal identities
$R^{13}R^{12}=(\id\o\Delta)(R)$, $R^{13}R^{23}=(\Delta\o\id)(R)$,
and the intertwining property $R\Delta(x)=\Delta^{op}(x)R,\
x\in \D(H)$, where $\Delta^{op}\colon x\mapsto x_{(2)}\o
x_{(1)}$.

If $\{b_A\}$ and
$\{\beta^A\}$ denote bases of $H$ and $\hat H$, respectively,
that are dual to each other, $\langle
\beta^A,b_B\rangle=\delta^A_B$, then
\beq
R\equiv R_1\o R_2:=\sum_A\ \D(b_A)\o\D(\beta^A)
\eeq
is independent of the choice of the bases and satisfies the
above identities.

An important
theorem proven by Drinfeld [Dr2] claims that in a
quasitriangular Hopf algebra $\G(m,u,S,\Delta,\varepsilon,R)$ there
exists a canonically chosen element $s\in \G$ implementing the
square of the antipode, namely $s=S(R_2)R_1$.
Its coproduct is related to the $R$-matrix by the
equation
$$\Delta(s)= (R^{op}R)^{-1}(s\o s)=(s\o
s)(R^{op}R)^{-1}\eqno(B.6)$$
which turns out to mean that $s$ defines a universal balancing
element in the category of representations of $\G$.

The universal balancing element $s$ of
$\D(H)$ takes the form
$$
s:= S_{\D}(R_2)R_1\equiv \D(S^{-1}(\beta^A))\D(b_A)
\eqno(B.7)
$$
and if $H$ (and therefore $\D(H)$) is a $C^*$-Hopf algebra then
$s$ is a central unitary of $\D(H)$. Its inverse can be written
simply as
$$s^{-1}=R_1R_2=R_2R_1\,.\eqno(B.8)$$
The existence of $s$ satisfying (B.6) is needed in Section 4.1 to prove
that in the Hopf spin model
the two-point amplimorphisms (and therefore, by Lemma
3.16, {\it all} universal amplimorphisms) are strictly
translation covariant.

\end{appendix}

\bsn

{\bf Acknowledgements:} F.N. would like to thank H.W. Wiesbrock
for stimulating interest and helpful discussions.

\end{document}